\documentclass[aps,prx,amsfonts,floatfix,superscriptaddress,twocolumn,longbibliography]{revtex4-2}
\usepackage{mathbbol}
\usepackage{graphicx,color,xcolor}
\usepackage{amsmath,amssymb,bm,amsthm,amstext}
\usepackage{amsfonts}
\usepackage{mathtools}
\usepackage{simplewick}
\usepackage{braket}
\usepackage{qcircuit}
\usepackage{array}
\usepackage{extarrows}

\usepackage{mathtools}
\DeclarePairedDelimiter\dbra{\langle\langle}{\rvert}
\DeclarePairedDelimiter\dket{\lvert}{\rangle\rangle}

\usepackage{soul}

\usepackage{xfrac}
\usepackage{tikz}
\usepackage{circuitikz}
\usepackage{epstopdf} 
\usepackage{times}
\usepackage{dsfont}
\usepackage[normalem]{ulem}

\usepackage{bm}
\usepackage{mathrsfs}

\DeclareMathAlphabet{\mathtensor}{OT1}{cmss}{sbc}{n}
\DeclareMathAlphabet{\bmathtensor}{OT1}{cmss}{bx}{n}

\newcommand{\e}{\text{e}}

\begin{document}

\title{Quantum Gates Between Mesoscopic Spin Ensembles}

\author{Mohamad Niknam}
\address{Department of Chemistry and Biochemistry, University of California Los Angeles, 607 Charles E. Young Drive East, Los Angeles, California 90095-1059, USA}
\address{Center for Quantum Science and Engineering, UCLA}
\author{Robert N. Schwartz}
\address{Department of Electrical and Computer Engineering, UCLA}
\author{Louis-S. Bouchard}
\address{Department of Chemistry and Biochemistry, University of California Los Angeles, 607 Charles E. Young Drive East, Los Angeles, California 90095-1059, USA}
\address{Center for Quantum Science and Engineering, UCLA}
\address{California NanoSystems Institute, UCLA}

\date{\today}

\begin{abstract}
Quantum algorithmics with single spins poses serious technological challenges  such as precision fabrication, rapid decoherence, atomic-scale addressing and readout. To circumvent atomic-scale challenges, we examine the case of fully polarized mesoscopic spin ensembles (spin-coherent states) whose total angular momenta states map to qudit submanifolds.  We show that in the limit where the size of  the ensembles is small compared to their separation, it is possible to treat them as qubits with an effective coupling strength that scales with the number of spins.  If the spins within each ensemble are decoupled (e.g., via control fields, spinning or diffusional averaging or materials engineering), one- and two-qubit gate operations can be implemented with high fidelities. 
\end{abstract}

\maketitle

{\em Introduction.--} 
Large-scale quantum operations are difficult to develop, requiring isolation of a large number of individual qubits, precise control of their quantum states, error correction schemes and high-accuracy measurements.  Existing qubits are implemented in  environments requiring high vacuums or low temperatures. Several recent initiatives have been driving hardware development for the realization of new quantum technologies that are more  scalable,  robust, and less physically demanding. Qubits should have long coherence times or be able to retain quantum information for times much longer than the operation time of quantum gates.  Arbitrary unitary transformations on the set of all qubits can be constructed up to desired precision by composing quantum gates chosen from a set of universal quantum gates.  The latter typically consists of single-qubit gates (one per qubit) and one two-qubit gate (acting on pairs of qubits)~\cite{Barenco95}.
Implementation of the two-qubit gate has proven to be the most challenging, and even the leading platforms for quantum computing, i.e., superconducting qubits~\cite{Martinis14}, trapped ions~\cite{Ballance16}, and Rydberg atoms~\cite{Bluvstein21} have error rates exceeding 0.1\% per gate, which is an order of magnitude larger than the threshold required for fault-tolerant quantum computing~\cite{Preskill18}.   
 
Ensemble quantum computing has the advantage of many replicas, eliminating the need to repeat projective readouts thousands of times to obtain the statistical weights in the many-qubit wavefunction.  Ensembles can have favorable properties such as ease of fabrication, built-in robustness, longer storage times, more sensitive detection, etc.   Liquid-state NMR is of historical significance because it provided  the platform for the first experimental demonstrations of a working quantum algorithm~\cite{Chuang97,Cory97}.  Nuclear spins in the liquid-state feature long coherence times, typically in the order of seconds.  They interact weakly, meaning that they behave independently and can be addressed individually.  Qubits can be selectively addressed using frequency-selective (soft) pulses. This allows coherent control of quantum devices containing up to 12 qubits~\cite{Negrevergne06} and implementation of Shor's quantum algorithm~\cite{Chuang01}. 
However, nuclear spin qubits are not scalable due to the difficulty of initializing pure quantum states (highly mixed pseudo-pure states above $\approx 1$ mK), and synthesizing molecules with a large number of individually addressable nuclear spins coupled to one another.  Indeed, molecules can only accommodate the selective addressing of a small number of distinct nuclei due to the narrow range of chemical shifts.  Therefore, obtaining entanglement in liquid-state qubits for ensemble NMR quantum computing is nearly impossible~\cite{Braunstein99}.   NMR ensemble computing involves performing collective quantum operations on molecules, followed by averaging the state of qubits over the ensemble.   
To address the important issues of scalability and ease-of-fabrication, we revisit the idea of ensemble computing and consider reversing the order of averaging: ensemble-average spins to obtain large qubits, then perform operations among the ensembles.  In doing so, we use the rotational properties of large-$J$ angular momentum operators, which leads to a realization of qubit transformations similar to the single spin case~\cite{bib:sanctuary1976}.    
  In this paper we will assume the ability to fully polarize spins and work with spin-coherent states that we term E-qubits.  Practical considerations are discussed.

For nuclear spins the notation $\mathbf{I}$ is used instead of $\mathbf{J}$ to denote the spin angular momentum. Suppose we have an ensemble of $N$ molecules, each with $n$ nuclear spins $\{ \mathbf{I}^1,\dots, \mathbf{I}^n \}$.  The Hilbert space for the nuclear spin degrees of freedom in a molecule has dimension $\prod_{i=1}^n (2I^i+1)$.  The state of each molecule's nuclear spins is represented by a projector $P_\omega(\mathbf{I}^1,\dots,\mathbf{I}^n)$, $\omega=1,\dots,N$.  Let $\ket{ \varphi_i(\omega) }$ be the wavefunction describing the state of spin $i$ on molecule $\omega$.  Each projector is initially a product state, i.e., $P_\omega = \ket{ \psi_\omega } \bra{ \psi_\omega }$, where $\ket{ \psi_\omega } = \ket{ \varphi_1(\omega) } \otimes \dots \otimes \ket{ \varphi_n(\omega) }$, i.e.,
$$  P_\omega = \ket{ \varphi_1(\omega)  } \bra{ \varphi_1(\omega)  } \otimes \dots \otimes  \ket{ \varphi_n(\omega)  } \bra{ \varphi_n(\omega)  }. $$
Here, $\omega$ is standard notation in probability theory to denote the elementary outcome of a random variable.
The unitary propagator corresponding to the quantum circuit is denoted $U(\mathbf{I}^1,\dots,\mathbf{I}^n)$.  The final projector is $U(\mathbf{I}^1,\dots,\mathbf{I}^n) P_\omega(\mathbf{I}^1,\dots,\mathbf{I}^n) U(\mathbf{I}^1,\dots,\mathbf{I}^n)^\dagger$.  Averaging over $N$ molecules (denoted by overline):
\begin{multline*}
 \overline{U(\mathbf{I}^1,\dots,\mathbf{I}^n) P_\omega(\mathbf{I}^1,\dots,\mathbf{I}^n) U(\mathbf{I}^1,\dots,\mathbf{I}^n)^\dagger} \\
  = U(\mathbf{I}^1,\dots,\mathbf{I}^n) \rho(\mathbf{I}^1,\dots,\mathbf{I}^n) U(\mathbf{I}^1,\dots,\mathbf{I}^n)^\dagger \equiv  \rho'
\end{multline*}
where 
\begin{multline*}
\rho(\mathbf{I}^1,\dots,\mathbf{I}^n) \equiv \overline{P_\omega(\mathbf{I}^1,\dots,\mathbf{I}^n)} = \sum_{\omega=1}^N p_\omega P_\omega(\mathbf{I}^1,\dots,\mathbf{I}^n)
\end{multline*}
is a statistical (density) operator and $\sum_\omega p_\omega=1$.  The readout of  observable operator $A$ is obtained as:
$$ A^{(e_1)} \equiv \braket{ A(\mathbf{I}^1,\dots,\mathbf{I}^n) } = \mbox{Tr}[ \rho'(\mathbf{I}^1,\dots,\mathbf{I}^n) A(\mathbf{I}^1,\dots,\mathbf{I}^n)]$$
where $e_1$ distinguishes this form of ensemble averaging. On the other hand, suppose we start with $n$ ensembles, each described by a statistical operator, $\rho_i(\mathbf{I}^i)$, describing this initial ensemble averaging.   Each ensemble contains $N$ spins. The initial state is a product state $\rho_1(\mathbf{I}^1) \otimes \dots \otimes \rho_n(\mathbf{I}^n)$, where
$$ \rho_i(\mathbf{I}_i) = \sum_\omega p_{\omega,i} \ket{ \varphi_i(\omega) }  \bra{ \varphi_i(\omega) }, \quad i=1,\dots,n. $$
The propagator for the quantum circuit is denoted $V(\mathbf{I}^1,\dots,\mathbf{I}^n)$.  (The notation $V$ instead of $U$ indicates that the circuit implementations may be different, depending on the particular physical implementation of the qubits.)  Evolution of this product state leads to: 
$$\rho''=V(\mathbf{I}^1,\dots,\mathbf{I}^n) \rho_1(\mathbf{I}^1) \otimes \dots \otimes \rho_n(\mathbf{I}^n) V(\mathbf{I}^1,\dots,\mathbf{I}^n)^\dagger$$
The readout for this ensemble-averaging method, labeled $e_2$, leads to:
$$ A^{(e_2)} \equiv \braket{ A(\mathbf{I}^1,\dots,\mathbf{I}^n) } = \mbox{Tr}[ \rho''(\mathbf{I}^1,\dots,\mathbf{I}^n) A(\mathbf{I}^1,\dots,\mathbf{I}^n)].$$
As a consequence of the different orders of averaging, we note that even if $U=V$, $A^{(e_1)}$ and $A^{(e_2)}$ do not generally give the same result.  The averaging to obtain $A^{(e_1)}$ only requires specifying the probability distribution $\{ p_\omega \}_{\omega=1}^N$ whereas averaging for $A^{(e_2)}$ is described by a set  $i=1,\dots, n$ of distributions $\{ p_{\omega,i} \}_{\omega=1}^N$, $\sum_\omega p_{\omega,i} =1$.

Averaging over molecules before any computation is expected to yield different outcomes.  Qubits occupying larger physical volumes would significantly relax the manufacturing requirements, as the placement of single-atom qubits in a solid lattice (or in a host molecule, for that matter) at precise locations and spacings has been challenging due to the lack of methods to implant point defects with good precision~\cite{Simmons12}. A second advantage is the robustness of the quantum operations with respect to environmental noise.  A single spin undergoing decoherence leads to the loss of its quantum state, whereas the quantum state stored in ensembles may survive longer due to the inherent robustness of ensembles (see Appendix~\ref{appA}). 

\begin{figure}[htb]
\centering
\includegraphics*[width=0.95\columnwidth]{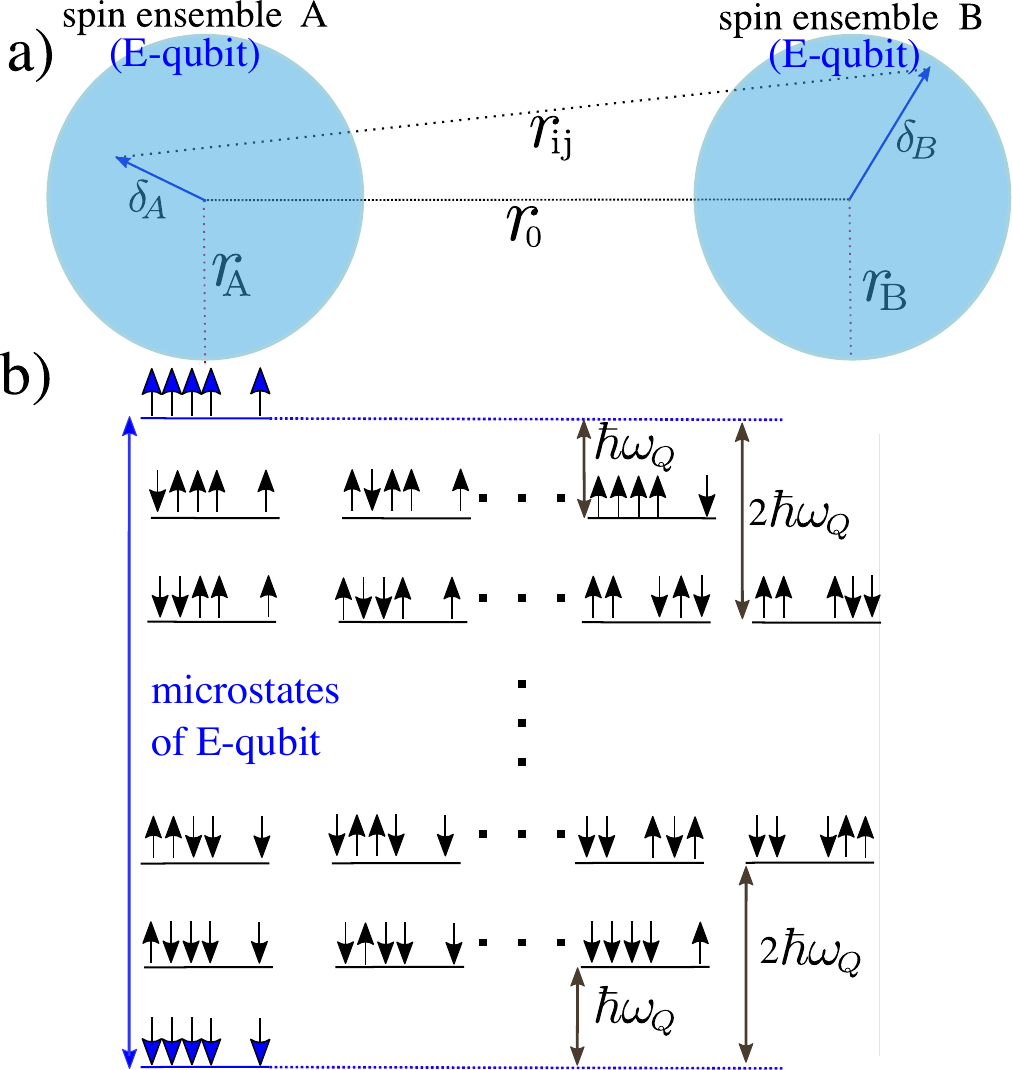}	   	 	 
\caption{a) Schematic of two E-qubits $A$ and $B$ coupled over mesoscopic distance. Here, $\vec{r}_{ij} = \vec{r}_0 + \vec{\delta}_B - \vec{\delta}_A \equiv \vec{r}_0 + \vec{\epsilon}$. For $\epsilon \ll 1$ the Hamiltonian of these two $n$-spins systems is equivalent to a two qubit Hamiltonian. We denote  magnitudes of vectors without an arrow, i.e., $\delta_A \equiv |\vec{\delta}_A|$, etc. b) Energy levels for an E-qubit.  The single-quantum transition energy is denoted $\hbar \omega_Q$.  The states with all spins up and all spins down, constitute the two level structure of a qubit. Control sequences are applied  on collective state of spins. The case with $I=1/2$ and single-spin basis states $\ket{\uparrow}$ and $\ket{\downarrow}$ is shown.}
\label{fig:Qubits}
\end{figure}

Suppose we have a large ensemble ($N\cdot n \gg 1$) of nuclear spins $\{\mathbf{I}^i\}$, where $N$ is the number of localized spin ensembles and $n$ is the number of spins per ensemble.  We will call these mesoscopic spin ensembles E-qubits for reasons that will become clear shortly. To simplify the presentation we assume that $I=1/2$ for nuclear spins;  we will also assume that all spins in each localized ensemble have the same gyromagnetic ratio. This choice only affects some constants and scaling factors below, and does not lead to a loss of generality. The magnetic dipole interaction between $N\cdot n$ spins cannot be used for quantum computing because of the impossibility of addressing each spin individually. Also, the exact couplings in atomic ensembles are not known, making it impossible to design precise gates. We consider: (1) two ($N=2$) localized ensembles $A$ and $B$ each containing $n$ spins; (2) assume full polarization of all spins in the ensembles; (3) spins within each ensemble are decoupled. Spin ensembles are rarely discussed as candidate qubits because of their multilevel energy structure and high degeneracies, which differs from the structure of a qubit. If each ensemble admits a $\mathfrak{su}(2)$-like algebra, the ensembles can be viewed as qubits, even though they are in reality multilevel qudits (Fig.~\ref{fig:Qubits}). The magnetic interaction between all pairs of spins is 
\begin{equation}
\! \! \sum_{i\in A, \atop j\in B} \mathbf{I}^i  \cdot \mathtensor{D}_{ij} \cdot \mathbf{I}^j, \; \; \mathtensor{D}_{ij} =-\frac{\mu_0}{4\pi}\frac{\gamma_i \gamma_j \hbar^2}{|\vec{r}_{ij}|^3} \left(3\frac{ \vec{r}_{ij} \otimes   \vec{r}_{ij}}{|\vec{r}_{ij}|^2}- \mathbb{1}_3 \right) 
\label{eq:Hq0}
\end{equation}
where $\vec{r}_{ij}$ is the vector connecting spin $i$ from $A$ to spin $j$ from $B$ (Fig.~\ref{fig:Qubits}a).  $\mu_0$ is the vacuum permeability and $\gamma$ is the gyromagnetic ratio. If there are $N$ such localized spin ensembles, each containing $n$ spin 1/2 particles the dimension of the spin algebra $\mathfrak{su}(2^{Nn})$ is $2^{2Nn}-1$. Clean quantum gates are still not possible due to the random distribution of $\vec{r}_{ij}$ values.  Moreover, this type of interaction over macroscopic distances is rarely considered due to the rapid $1/r^3$ falloff of the dipolar interaction.  However, in a specific cooperative geometry, clean gates are possible via cooperative couplings, as explained below.

Consider a geometry involving two spherical volumes $A$ and $B$ (Fig.~\ref{fig:Qubits}a) whose centers are separated by a constant vector $\vec{r}_0$ and the distance of each spin from the center of each region is much smaller than $|\vec{r}_0|$, i.e., $\delta_A \ll r_0$, $\delta_B \ll r_0$. The internuclear vector $\vec{r}_{ij}$ is nearly equal to $\vec{r}_0$. Expanding $|\vec{r}_{ij}|^{-3}$ in the small parameter $\vec{\epsilon}=\vec{\delta}_B-\vec{\delta}_A$ about $\vec{r}_0$ gives
\begin{equation}
\frac{1}{|\vec{r}_0+\vec{\epsilon}|^3}=\frac{1}{r_0^3} -  3 \frac{\vec{r}_{0} \cdot \vec{\epsilon} }{r_0^5} 
+ O(|\epsilon|^2).
\label{eq:taylor}
\end{equation}
The zeroth order term (i.e., neglecting terms of order $|\epsilon/r_0|$ and higher) gives the effective Hamiltonian $\mathcal{H}_D$
$$\sum_{i\in A, \atop  j\in B} \mathbf{I}^i \cdot \mathtensor{D}_{AB} \cdot \mathbf{I}^j, \quad 
\mathtensor{D}_{AB}=-\frac{ \mu_0 \gamma_A \gamma_B \hbar^2 }{ 4\pi r_0^3} \left( 3 \frac{ \vec{r}_0 \otimes \vec{r}_0 }{ r_0^2 } - \mathbb{1}_3 \right),$$ 
where the coupling tensor is now constant. It is therefore independent of $i$ and $j$, the indices of summation.    The Hamiltonian can be written as:
\begin{equation}
\mathcal{H}_D=\bigl( \sum_{i=1}^n \mathbf{I}^{A,i} \bigr)  \cdot  \mathtensor{D}_{AB} \cdot \bigl( \sum_{j=1}^n \mathbf{I}^{B,j} \bigr) =\mathbf{I}^A \cdot \mathtensor{D}_{AB} \cdot \mathbf{I}^B, \label{eq:newH}
\end{equation}
where $\mathbf{I}^A=\sum_{i=1}^{n} \mathbf{I}^{A,i}$ and $\mathbf{I}^B = \sum_{j=1}^{n} \mathbf{I}^{B,j}$ are total angular momentum  operators.   We thus have a bilinear coupling $\mathbf{I}^A \cdot \mathtensor{D}_{AB} \cdot \mathbf{I}^B$ involving two large  spins $A$ and $B$. 

Also, since $[I_{\alpha}^A,I_{\beta}^A] = i\epsilon_{\alpha\beta\gamma}I_{\gamma}^A$, $ [\mathbf{I}^A,\mathbf{I}^B] = 0$, the action of the total angular momentum operator on the composite $\prod_{i=1}^n (2I^i+1)$ dimensional Hilbert space of a localized ensemble constitutes a representation of the $\mathfrak{su}(2)$ Lie algebra that is reducible. Due to this rotational symmetry that maps to that of qubits, ensembles $A$ and $B$ are examples of E-qubits. (The $\mathfrak{su}(2)$ symmetry of large spins has been known since 1932~\cite{bib:majorana1932}.) If we have $N$ localized spin ensembles, each with $n$ spins, the  spin-Hamiltonian is the sum of Zeeman (static, control fields) and point dipolar interaction between the E-qubits, \begin{align}
\mathcal{H}_Z + \mathcal{H}_D = \sum_{Q=1}^N \mathbf{I}^Q \cdot \bigl(\vec{h}^Q(t) + \frac{1}{2} \sum_{P \ne Q}^N \mathtensor{D}_{QP} \cdot \mathbf{I}^P 
\bigr)
\label{eq:Hqd}
\end{align}
with $\vec{h}^Q=(h_{x}^Q,h_{y}^Q,h_z^Q)$ is an external field. It
has $\otimes \mathfrak{su}(2)^N \simeq \mathfrak{su}(2^N)$ symmetry.  Even though E-qubits are multilevel systems, the Hamiltonian~(\ref{eq:Hqd}) of the $nN$-spins system is equivalent to an effective $N$-qubit Hamiltonian, which is controllable. Controllability would not be possible with the lower symmetric  form~(\ref{eq:Hq0}).  The spin operator algebra is also much larger, with dimension $(2I+1)^{2Nn}-1$ for ~(\ref{eq:Hq0}) vs $(2I+1)^{2N}-1$ for Eq.~(\ref{eq:Hqd}). 

{\em Qubit state.--} Each E-qubit occupies a mesoscopic volume and can therefore be addressed selectively using nanomagnets~\cite{niknam22_nanomagnets}, nanowires, or similar technologies. Assuming $I=1/2$, the density matrix of subsystem $A$ containing $n$ fully polarized spins $(\frac{1}{2} \mathbb{1}\pm I_z)^{\otimes n}$, corresponding to $\ket{I^A,\pm I^A} \bra{I^A,\pm I^A}$, in product operator notation is
$$\frac{1}{2^n}\sum_{m=0}^n 2^{m-n} (\pm 1)^m \mathbb{1}_{2^n}^A \hspace{-0.15in} \sum_{i_1 < \dots < i_m}  \hspace{-0.15in} I_{z}^{A,i_1} I_{z}^{A,i_2} \dots I_{z}^{A,i_m} $$
\noindent where $$ I_{z}^{A,i} =\overbrace{ \mathbb{1}_2^A \otimes \mathbb{1}_2^A \otimes \dots \underbrace{ \otimes (\sigma_z^A/2) \otimes }_{i-\mathrm{th~position}} \dots \otimes \mathbb{1}_2^A \otimes \mathbb{1}_2^A}^{n ~\mathrm{factors}}. $$
A collective $\pi$ rotation on all $n$ spins using the control field $h_x^A(t) I_x^A \otimes \mathds{1}^B$, reverses the sign of all $m$-odd terms, flipping $\ket{I^A, I^A}\bra{I^A, I^A}$ into $\ket{I^A,- I^A}\bra{I^A,- I^A}$ and vice versa. These states  map to the usual computational basis $\ket{0}_A$ and $\ket{1}_A$. Here we do not have selectivity to address intermediate transitions within each ensemble, so these intermediate levels are not accessible through unitaries  (Fig.~\ref{fig:Qubits}) or decoherence (see Appendix~\ref{appA} and~\ref{appA2}).

The microstates of an E-qubit can be viewed as a very large $d$-level system. For a $d$-level system a natural quantum computational basis consists of the top and bottom energy levels. Transitions between the top and bottom levels of a $d$-level system are multiple-quantum transitions requiring frequencies proportional to $d$. When $d$ is large, such energies become out of reach to the experimentalist. On the other hand, we know from decades of NMR and EPR experiments that very large $d$ (multi-quantum) transitions are never needed if the spins are uncoupled or weakly coupled (i.e., interaction strength is much smaller than the Zeeman energy). In the weak coupling limit, only single-quantum excitations are needed, as each spin can be viewed as behaving independently. In this case, unitary transformation rules of conventional quantum computing with qubits (spin 1/2) or those of qudits (see Appendices~\ref{appB} and~\ref{appC}) for large spins apply, as far as the generation of single- and two-qubit gates. If the spins were coupled sufficiently strongly as to create a large spin, impractically high energy excitations may be needed to excite magnetic transitions between the lowest and highest energy states.  To reduce the ion-ion couplings engineered materials may be required in which the sum of all interactions (e.g., dipolar, indirect spin-spin, or exchange between neighboring spins is weak. A notable approach has been magnon-cavity coupling where introduction of strong exchange coupling results in narrow-linewidth magnetostatic modes~\cite{tabuchi2015coherent,lachance2017resolving}.

\begin{figure}[htb]
\centering
\includegraphics*[width=0.95\columnwidth]{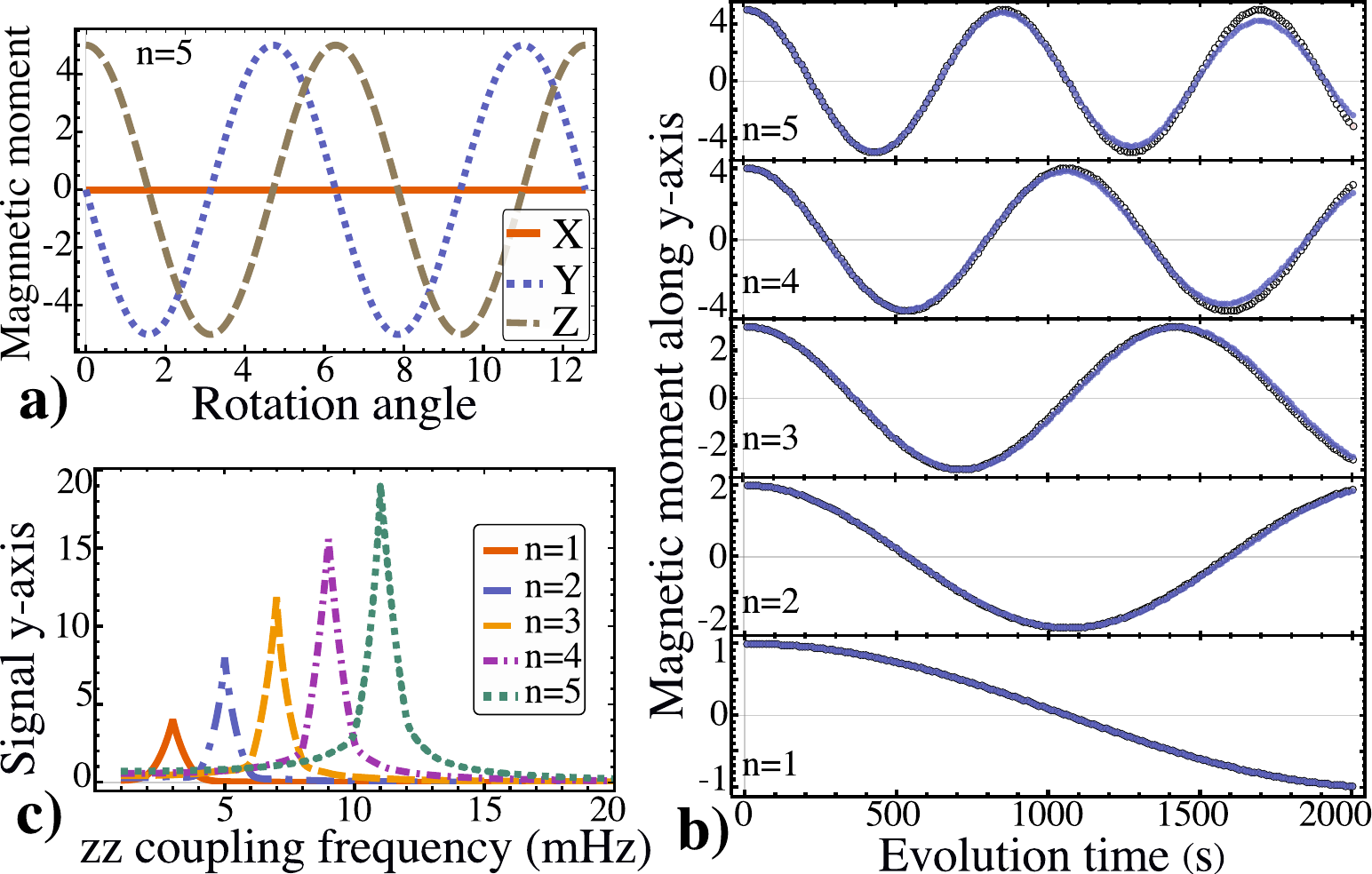}	   	 	 
\caption{Demonstration of single- and two-qubit gates for E-qubits. Single qubit gates are applied with standard rotation operators; the control frequencies are independent of $n$. a) Rabi oscillations for the $n=5$ with $R_x^A(\theta)$ applied to E-qubit A. The vertical axis shows the amplitude of observed signal resulting from projection of the density matrix on the $X$, $Y$, and $Z$ axes. Spins are chosen at random locations inside two spheres with $r_A=r_B=30$ nm and $r_0=100$ nm.  b) Evolution under the Ising $zz$ interaction needed for the  entangling CNOT gate, with $ n=1-5$. Black/blue dots project the evolution of the E-qubit B with full/effective Hamiltonian, Eq.~\eqref{eq:Hq0}/Eq.~\eqref{eq:newH}, on the vertical axis. c) Fourier transform of signal from the E-qubit B evolving with the full Hamiltonian, blue points  from panel b, show a clear linear dependency of the coupling frequency to $n$. The signal amplitude also grows with $n$ as expected.
} 	
\label{fig:gates}
\end{figure}

{\em Quantum gates.--} To have universal quantum gates we need arbitrary single qubit gates complemented by one entangling two-qubit gate. Suppose the initial state of an E-qubit is fully polarized, e.g., $\ket{I^A,\pm I^A}$. This fully polarized state corresponds to a linear combination of fully polarized high-rank axial tensor operators $\mathscr{Y}^{(k)0}(\mathbf{I}^A)$, $0 \le k \le 2I^A$, such that the single-qubit rotations are described by  $\sum_q \mathscr{D}^{(k)}_{q0}(\Omega) \mathscr{Y}^{(k)0}(\mathbf{I}^A)$, where $\mathscr{D}^{(k)}_{qq'}(\Omega)$ are elements of Wigner $D$ matrices representing the rotation group in the subspace of E-qubit $A$ (see Appendices~\ref{appB} and~\ref{appC}).  These collective rotations can be implemented using unitary operators of the form $\e^{-i \theta \hat{n} \cdot \mathbf{I}^A} \otimes \mathbb{1}^B$, where $\hat{n}$ is a rotation axis and $\theta$ is the angle of rotation.  Transformations in the full rotation group generate coherent rotations needed for single-qubit gates.  This will work as long as the spins within each ensemble are decoupled from each other.  The fidelity of quantum gates depends on linewidth. The latter is influenced by couplings between spins in each E-qubit. The fidelity of a quantum gate, to the first order, depends linearly on relaxation rate and inverse of gate time~\cite{abad2022universal}.
 We discuss decoupling schemes below.  Figure~\ref{fig:gates}(a) demonstrates such rotations of spin-coherent states describing the E-qubit with $n=5$ and $N=2$.   Initially polarized states such as $ \ket{I^A, I^A} $ in the coupled spin basis can easily be rotated into (say) $\ket{I^A,-I^A}$ using standard operations.

The CNOT gate where the first E-qubit  is the control and the second  is the target can be implemented with the following sequence of unitaries:
$$\sqrt{i} R^A_z(\frac{\pi}{2}) \cdot R^B_z(\frac{-\pi}{2}) \cdot R^B_x(\frac{\pi}{2}) \cdot U_{zz}(\frac{1}{2\|\mathtensor{D}_{AB}\|}) \cdot R^B_y(\frac{\pi}{2}), $$ 
where $\| \mathtensor{D}_{AB}\|$ is the norm of $\mathtensor{D}_{AB}$, $U_{zz}(\frac{1}{2\|\mathtensor{D}_{AB}\|})=\exp(i \pi I_z^A \tiny{\otimes} I_z^B)$ describes evolution under an Ising ($zz$)-type interaction derived from Eq.~(\ref{eq:Hqd}). Figure~\ref{fig:gates}(c) indicates simulated time-evolution under the full $U_{zz}$ dynamics for two qudits $N=2$, with $n=1,2,3,4,5$ spins, showing that the behavior is similar to the evolution of  qubits, except that the evolution frequency scales with the number of spins in each ensemble. 
 Figure~\ref{fig:gates} proves that the coupling between E-qubits scales linearly with $n$.  Although cases $n> 5$ ($>$ 10 spins total) are difficult to simulate on an average classical computer, this linear relationship holds for any $n$.  We discuss below specific examples for much larger $n$. With spin ensembles, storage of quantum information is inherently robust to noise if the spin-flips affect only a small fraction of the total moment. This is an advantage over single-atom qubits. We now discuss decoupling schemes.

{\em Solid-state nuclear spins.--} 
In solids, the magnetic dipole interaction between nearest neighbors is strong. In contrast to the liquid state interactions, the spatial part of dipolar coupling, $(1-3\cos^2\theta_{ij}) r_{ij}^{-3}$, is not averaged to zero as the result of  molecular motion. Spin evolution under homonuclear dipolar interaction  $\sum_{i<j} D_{ij} (2I_{z}^{i}I_{z}^{j}-\frac{1}{2}(I_{+}^{i}I_{-}^{j}+I_{-}^{i} I_{+}^{j}))$ transforms uncorrelated spin terms such as $I_{x}^{i} \otimes \mathds{1}^j$   into correlated terms such as $I_{y}^{i} \otimes I_{z}^{j}$. By connecting multiple spin pairs, a network of multispin correlated terms is generated that grows in size exponentially fast. These correlated terms are not directly observable and result in short $T_2$ times, on the order of microseconds~\cite{Munowitz87,Cho2005,Niknam20,Niknam21}. 

\begin{figure}[htb]
\centering
\includegraphics*[scale=0.5]{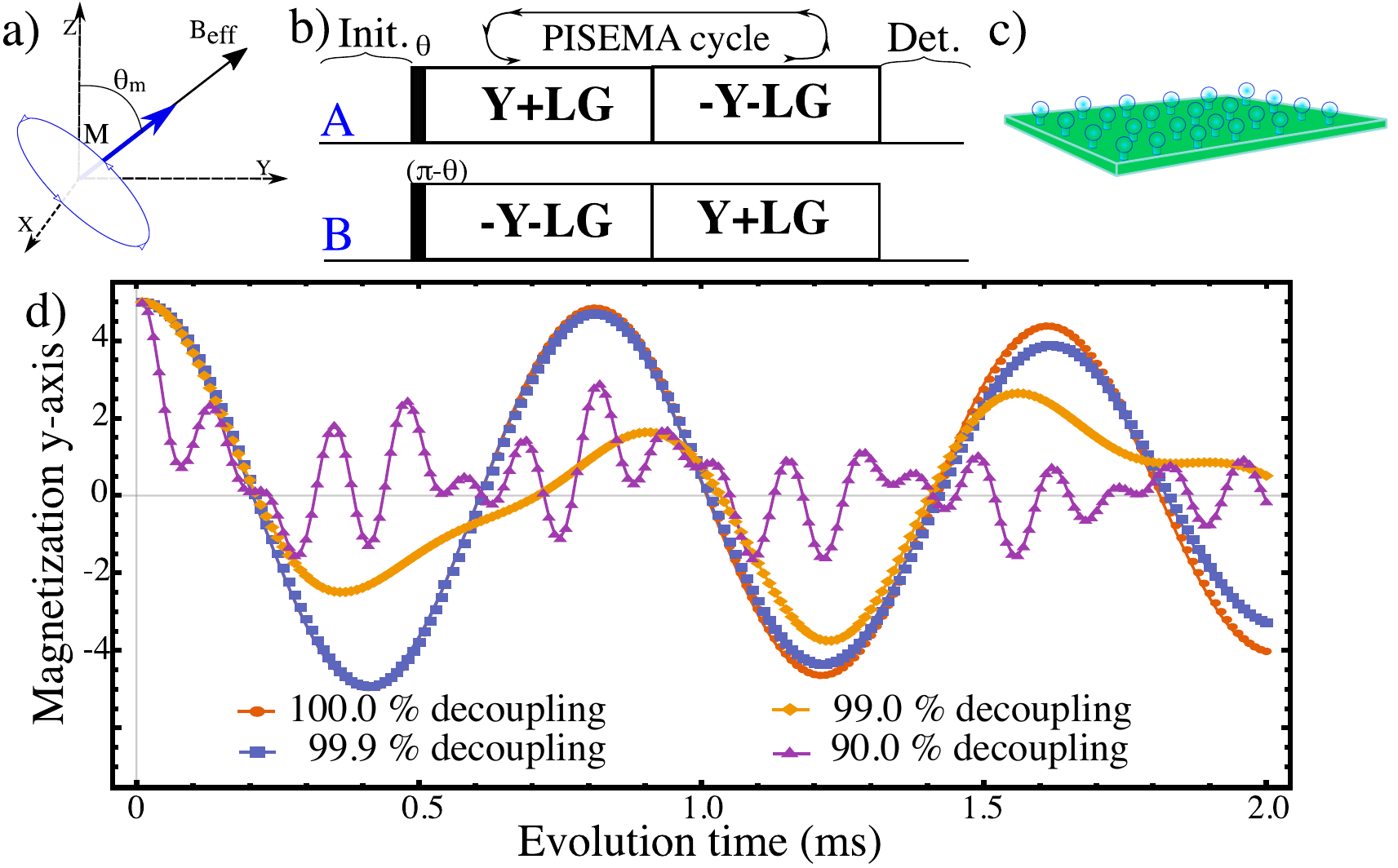}
\caption{Quantum operations with solid state nuclear-spin ensemble qubits. a) In the LG spin-locking experiment, an effective field along the magic angle is applied to remove the dipolar interaction inside each E-qubit. b)  Decoupling sequence for removing the intra-qudit dipolar interaction in E-qubits $A$ and $B$, while preserving the $zz$ interaction between them. c) Proposed experimental setup for nuclear spins. The radius of each sphere is 30~nm and they are 100~nm apart. d) Two-qubit entangling gate between two E-qubits.  This experiment is limited by the efficiency of control sequence in removing the intra-qubit interaction.  Residual interactions larger than $1\%$ have a devastating effect on the fidelity of the two-qubit gate.  Average gate fidelity reduces linearly with the linewidth of spins in E-qubits.
} 	
\label{fig:solidqubit}
\end{figure}

To average out the dipolar interactions inside each E-qubit while preserving the inter-qubit interactions, well-known techniques of dynamic decoupling can be applied to average out unwanted dipolar interactions and obtain the desired effective Hamiltonian. The idea is to either suppress unwanted interactions or shift them to a different frequency band. One may leverage solid-echo cycles such as MREV-8~\cite{Rhim73_mrev8}, or sequences based on magic echo~\cite{Boutis03}, or use magic angle spin-locking such as Lee-Goldberg (LG)~\cite{Lee65}. Consider the  polarization-inversion spin-exchange at the magic angle (PISEMA) experiment which is a variant of LG decoupling sequence.  The homonuclear dipolar interaction is averaged to zero by locking the spins  to an effective field $\mathbf{B}_{\text{eff}}$ along the magic angle, $ \theta_{m}=\arccos(\sqrt{1/3})$. The spatial part of the dipolar interaction averages to zero in each E-qubit, as long as $\| \mathbf{B}_{\text{eff}}\| \gg \|\omega_D\|$ (Fig.~\ref{fig:solidqubit}a). To understand the influence on the inter-qubit interaction, consider the magnetic moment of spins in each qubit. As their magnetic moment vectors rotate around $\mathbf{B}_{\text{eff}}$, the perpendicular component gains a time dependence and averages out over the cycle, while the parallel component survives. For all the spins with initial magnetic moment along the $z$ axis, a portion of their initial magnetic moment remains after the application of the locking field $ \tilde{I}_{z}^{\|,i}=I_{z}^{i} \sin{\theta_{m}}=0.82 \times I_{z}^{i}$ ~\cite{Pisema99}. This is a large scaling factor in comparison to rival decoupling sequences. Figure~\ref{fig:solidqubit}b shows the pulse sequence that combines these ideas for quantum control of two E-qubits. By applying the PISEMA sequence on both E-qubits simultaneously, we ensure that the intra-qubit interactions are averaged to zero. Selectivity in applying $\mathbf{B}_{\text{eff}}$ (direction and amplitude) means that the effective inter-qubit interactions can be engineered.  In particular, the effective magnetic moment in each qubit can be oriented at a different direction and with different precession frequencies resulting in the inter-qubit interaction of the form $\sin{(\theta_m)}^2\sum_i I_{z}^{A,i} \otimes \sum_j I_{z}^{B,j}$. 

To get an idea of the effective coupling strength, consider the example of solid hydrogen  ($T<14.01$~K, density $0.086$ g/cm$^3$). The gyromagnetic ratio of $^1$H is $26.75 \times 10^7$~rad/T/s. Consider an arrangement of spin ensembles each confined to spherical volumes with radius of $r=30$~nm. The centers of the ensembles are placed $r_0=100$~nm away from the centers of their nearest neighbors. There are $5.9\times 10^6$ spins in a sphere of solid hydrogen, leading to $D_{AB}=7.51 \times 10^{-4}$~rad/s. The magnetic field seen by a spin in sphere $A$ due to the field of spins in sphere $B$ is $104~\mu$T.  This corresponds to a precession frequency of 4.4 kHz. Entangling gates between the two ensembles can therefore be executed in 227~$\mu$s. This proposed implementation is limited by the efficiency of the control sequence in removing intra-qubit interactions. The fidelity of two-qubit  gate is mainly determined by the robustness of two-qubit unitary interactions such as $zz$. Figure~\ref{fig:solidqubit}d shows the effect of residual intra-qubit  interaction in the implementation of a macroscopic Ising $zz$ interaction.


{\em Liquid state or gas phase.--}
In the gas or liquid state Eq.~(\ref{eq:Hqd}) automatically takes the form of a mesoscopic Ising ($zz$)-type interaction. In high magnetic fields, dipolar interactions between nearby molecules vanish due to diffusion, i.e., $\braket{3\cos^2\theta-1}_{S^2}=0$, leading to intra-qubit decoupling.  However, beyond the diffusion length, the magnetic dipole interaction does not vanish, leading to a nonzero mesoscopic inter-qubit dipolar interaction. The end result is an Ising $zz$ form $\mathcal{H}_{AB} \propto I^A_z I^B_z$~\cite{Warren96,Richter2000,Jeener2000,Tang04}, where the proportionality constant depends only on the time-averaged internuclear vector connecting pairs of spins between each ensemble.  In spite of molecular motions this effective mesoscopic long-range interaction is static in the geometry described here.  For spherical volumes ($r \sim 30$~nm) filled with  liquid hydrogen, $r$ is less than the diffusion length of H$_2$ molecules during the timescale of a NMR experiment, the dipolar interaction is averaged out within each sphere. The effective coupling strength for inter-qubit interactions when $r_0=100$ nm is 3.6 kHz. This is to be contrasted with the long relaxation times ($>$ 1~s) of nuclear spins in liquids. 

{\em Nuclear-spin polarization.--} In the initialization step all spins must be fully polarized, which is a non-trivial task for nuclear spins. Depending on the physical implementation of the E-qubits, hyperpolarization techniques have been developed which may be applicable including dynamic nuclear polarization~\cite{Maly08}, spin exchange optical pumping~\cite{Patange14} and parahydrogen induced polarization~\cite{Haake96}. We also note that nuclear ferromagnetic ordering has been observed at nanokelvin temperatures~\cite{bib:nfo1,bib:nfo2,bib:nfo3,bib:nfo4,bib:nfo5}.

{\em Molecular magnets.--}  Moving on from nuclear spins to discussing electronic spins we switch our notation from $\mathbf{I}$ to $\mathbf{S}$.
Molecular magnets are high-spin clusters in a generally anisotropic crystalline structure~\cite{MolecularM2000}.  They are modeled as giant spins~\cite{Gaita19}; Mn$_{25}$ has $S=\tfrac{51}{2}$~\cite{Mn2508} whereas Chen et al.~\cite{bib:chen2018} reported a giant spin ground state of magnitude $S=91$. A $T_2$ time of 31 ms was recently demonstrated for $^{171}$Yb$^{3+}$ at 1.2 K~\cite{bib:faraon}. They are possible building blocks for the physical implementation of quantum processors. A realization of Grover's quantum search algorithm was achieved by coherent manipulation of a TbPc$_2$ molecular magnet~\cite{Godfrin17}.  Unlike nuclear spins, nearly full thermal polarization of electron spins, $\langle S_z \rangle =  \mbox{Tr}( S_z^\dagger   e^{-\beta \mathcal{H}})/\mbox{Tr}(e^{-\beta \mathcal{H}}) \approx |S|$, $\mathcal{H} = D [S_z^2 - \tfrac{1}{3} S(S+1)] + E[S_x^2 - S_y^2] - g \mu_B \mathbf{S \cdot B}$, can be asymptotically approached at high magnetic fields and low temperatures.  As an example, Gd$^{3+}$ ions in GdW$_{30}$ complexes~\cite{bib:jenkins} ($D=1281$ MHz, $E=294$ MHz, $S=\tfrac{7}{2}$, $g=2$) can reach 95\% polarization at 3 T and 2 K. The crystal-field (CF) term is sufficiently weak (for GdW$_{30}$, 3\% of the Zeeman term) and does not interfere significantly with the spin dynamics. Weak CF interactions are a general feature of rare-earth ions, stemming from the isolated nature of the $4f$ electrons, making it possible to treat them like nearly free paramagnetic ions (unlike transition-metal ions whose CF splittings are much larger~\cite{bib:boca}). Other methods could be considered, such as chirality-induced spin selectivity (CISS), whose helical electrons are fully polarized at high temperatures~\cite{CISS20}.  

The methods of the previous section can be extended to ensembles of large spins ($S>1/2$) provided that the algebra of large spins is used.  Molecular magnets are normally treated as multipole tensors, $\mathscr{Y}^{(k)q}(\mathbf{S})$ where $0 \le k \le 2S$, $-k \le q \le k$, in the EPR literature~\cite{bib:boca}.  A full quantum treatment of multipoles (e.g., Ref.~\cite{bib:sanctuary1976}) is presented in Appendices~\ref{appB} and~\ref{appC}.  There, we demonstrate the creation of one- and two-qubit gates involving arbitrary spins.  In this context, it is worth reviewing the well-known relationship from angular momentum theory between coupled and uncoupled representations.  It is the basis that gives rise to large composite electronic spins.  A molecular magnet is made up of $n$ magnetic atoms each with spin $\{\mathbf{s}^i \}$ ($i=1,\dots,n$). The total spin is $\mathbf{S}=\sum_{i=1}^n \mathbf{s}^i$.  The Zeeman interaction with an  applied magnetic field $\mathbf{B}(t)$ is $\mu_B \mathbf{B}(t) \cdot  \sum_{i=1}^n \mathtensor{g}_i \cdot \mathbf{s}^i$, where $\mu_B$ is the Bohr magneton and $\{ \mathtensor{g}_i \}$ are $g$-tensors. Using the Wigner-Eckart theorem
$$ \braket{ \alpha' , j' m' | \mathscr{Y}^{(k)q} | \alpha,jm} = \braket{ jk;mq|jk;j'm'} \frac{ \braket{\alpha' j' || \mathscr{Y}^{(k)} || \alpha j } }{ \sqrt{ 2j'+1}} $$
applied to vectors:
$$ \braket{ \alpha' , j' m' | V^{(1)q} | \alpha,jm} = \braket{ jm' | J^{(1)q} | jm } \frac{ \braket{\alpha' j' | \mathbf{J \cdot V} | \alpha j } }{ \hbar^2 j(j+1) }, $$
where $\mathscr{Y}^{(k)q}$ are components of an irreducible tensor $\mathscr{Y}^{(k)}$, $\braket{ jk;mq|jk;j'm'}$ is a Clebsch-Gordan coefficient and $\braket{\alpha' j' || \mathscr{Y}^{(k)} || \alpha j }$ is a reduced matrix element, it is customary to  write this interaction as $\mu_B \mathbf{B}(t) \cdot \mathtensor{g} \cdot\mathbf{S}$. Here, $\mathbf{S}$ is the total spin of the molecular magnet and $\mathtensor{g}=\sum_i \mathtensor{g}_i c_i$, with $c_i = \braket{\alpha' S | \mathbf{S \cdot s}^i | \alpha S}/\hbar^2 S(S+1)$. 
This is certainly the case when only the multiplet with maximum spin $S$ is populated. Inside this multiplet, application of this theorem for $\mathbf{V}=\mathbf{s}^i$, $\mathbf{J}=\mathbf{S}$, $\ket{jm} \equiv \ket{SM}$, enables replacing  $\mathbf{s}^i$ by $c_i \mathbf{S}$, where $c_i$ is a numerical constant, i.e., $\pi \mathbf{s}^i \pi=c_i \pi \mathbf{S} \pi$, where $\pi$ is the projection operator onto the multiplet of highest spin allowed in the ground state of the molecular magnet, i.e. $\pi = \sum_{M=-S}^S \ket{SM}\bra{SM}$.  The numerical constants $c_i$ are determined by the nature of the exchange interactions (e.g., ferro- vs antiferromagnetic) between the constituent spins.   



{\em Solid State Electronic Spins.--}  Solid state systems such as rare-earth ion or transition-metal-doped semiconductors could be a potential platform for E-qubits.  In the dilute limit, spins are weakly interacting and residual intra-qubit couplings can be addressed using decoupling schemes as discussed earlier.  In the dense limit, however, ions interact strongly.  Fortunately, uncoupling may be possible with the use of co-dopants or materials engineering.  Long decoherence times of Er$^{3+}$ ions in the presence of oxygen co-dopant have been observed in the system Er-Si~\cite{hu2022single,ranvcic2018coherence,yin2013optical}; we note that the system Er:YSO is also of interest.  For Er$^{3+}$ the ZFS is smaller than the Zeeman splitting from Tesla fields, making it possible to polarize electron spins above $>$ 90\%.  Weak ZFS are a general feature of rare-earth ions.

{\em Conclusion.--}
The design requirements of quantum processors based on large collective spins (ensembles, molecules) may be less stringent that those based on individual atomic spins.  A scalable implementation of the latter would require precise placement of atomic defects in the lattice with atomic resolution, something that is not currently feasible at this time using foundry techniques.  Also, once the single atom decays, the entire quantum state is lost.  That is not the case for ensembles (E-qubits), where survival of the fittest prevails, so-to-speak.  Moreover, E-qubits can be selectively addressed through spatial degrees of freedom, rather than frequency (spectral) selectivity, providing a sensible path towards scalability.   To date, a prescription of gate implementation between collective spins has been lacking. Herein we examined the feasibility of using direct macroscopic magnetic dipolar interactions for the implementation of universal gates between E-qubits. We have shown how to create coherent evolution where each E-qubit behaves isomorphically to $\mathfrak{su}(2)$.  Under conditions of intra-qubit decoupling and effective mesoscopic Ising $zz$ interactions, possible implementations could include nuclear-spin ensembles in the solid, liquid and gas phases, as well as giant electronic spins from molecular magnets (including possibly, ensembles of molecular magnets or rare-earth ion dopants).  Previous implementations of ensemble qubits such as neutral atoms rely on interactions with an external field to apply quantum gates, where the coupling scales with $\sqrt n$~\cite{rabl2006hybrid,saffman2016quantum} as compared to $n$, a clear advantage in our proposal.  High-fidelity control and readout methods are of course, critical to the success of such a proposal.  To this end, we note the possibility of electric-field control~\cite{bib:electricsmm},  nanomagnet control~\cite{niknam22_nanomagnets}, nanowire control~\cite{bib:itohefield} and nanosquid readout~\cite{bib:nanosquid}.

\section{Appendix}

\subsection{Quantum Treatment of Single- and Two-Qubit Gates For Arbitrary Spin\label{appB}}

Here we give a full quantum treatment of gates for maximally polarized states involving single-spin states from arbitrary spins.   The formalism of state multipoles $\mathscr{Y}^{(k)q}(\mathbf{S})$~\cite{bib:sanctuary1976} is employed.   State multipoles are defined in terms of the angular momentum states $\ket{SM}$ by Eq.~(\ref{eq:TmatrixEl}) below.  Sets of arbitrary spins are best treated using the multispin formalism of Sanctuary~\cite{bib:sanctuary1976}. The multispin tensors for $N$ spins, $T^{(k)q}_{ \{ K \} }(k_1,k_2,\dots,k_N)$, are constructed from the $\mathscr{Y}$'s as follows:
\begin{multline}
 T^{(k)q}_{ \{ K \} }(k_1,k_2,\dots,k_N) = \left[ \prod_{i=1}^N (S^i)^{-1/2} \right]  \\
 \times  \sum_{q_1, \dots, q_N} \braket{ k_1 q_1 k_2 q_2 \dots k_N q_N | (k_1 k_2 \dots k_N) \{K \} kq } \\
\times \mathscr{Y}^{(k_1)q_1}(\mathbf{S}^1)  \otimes \mathscr{Y}^{(k_2)q_2}(\mathbf{S}^2) \otimes \dots \otimes \mathscr{Y}^{(k_N)q_N}(\mathbf{S}^N),
\label{eq:multis}
\end{multline}
where  $\braket{ k_1 q_1 k_2 q_2 \dots k_N q_N | (k_1 k_2 \dots k_N) \{K \} kq }$ is a generalized Clebsch-Gordan coefficient~\cite{bib:sanctuary1976}, 
$(S^i)=(2S^i+1)$ and $\{ K \} = K_1, K_2, \dots, K_{N-2}$ is the angular momentum coupling scheme (set of intermediate values). The adjoint is:
\begin{equation}
T^{(k)q}_{ \{ K \} }(k_1,k_2,\dots,k_N)^\dagger = (-1)^{k-q} T^{(k)-q}_{ \{ K \} }(k_1,k_2,\dots,k_N).
\label{eq:adj}
\end{equation}
These operators are orthonormal in the following sense:
\begin{multline}
 \mbox{Tr}\left[ T^{(k)q}_{ \{ K \} }(k_1,k_2,\dots,k_N)^\dagger T^{(k')q'}_{ \{ K' \} }(k_1',k_2',\dots,k_N') \right] \\
 = \delta_{kk'} \delta_{qq'} \delta_{ \{K\}, \{ K'\} } \prod_{i=1}^N \delta_{k_i, k_i'},
 \label{eq:orth}
\end{multline}
where $ \delta_{ \{K\}, \{ K'\} }  = \delta_{K_1,K_1'} \delta_{K_2,K_2'} \dots \delta_{K_N,K_N'}$.
In this formalism a scalar operator on the $n$ spins can be written $\phi=\sum_{k,\{ V\} } \phi^{(k)}_{ \{ V \} } \odot^k T^{(k)}_{ \{ V \} }$, where $T^{(k)}_{ \{ V \} }$ forms a basis for the irreducible representation of the rotation group, $\phi^{(k)}_{ \{ V \} }$ are $k$-th rank tensor coefficients and $\odot^k$ indicates a $k$-th order contraction of the two tensors~\cite{bib:sanctuary1976}.  In this formalism the rotational invariance can be exploited but not when the expansion $\phi = \sum_{\alpha, \beta} \ket{\alpha} \bra{\alpha} \phi \ket{\beta} \bra{\beta}$ is used.  Here the basis states $\ket{\alpha}$, which form a complete set, are usually taken to be the product state of the $n$ spins.  Such a treatment with product states can be quite involved because the whole series must be used~\cite{bib:sanctuary1976}.

\subsubsection{Single-Qubit Gates}

It is clear by inspection of Eq.~(\ref{eq:multis}) that single-spin operators ($k=1$, $k_i=1$, $k_j=0$, $j \ne i$ for some $i=1,\dots,N$), i.e.,
$$  T^{(1)q}_{ \{ K \} }(00 \dots 1_i \dots 00) \propto \mathscr{Y}^{(1)q}(\mathbf{S}^i) \propto \mathbf{S}^i $$
in the spherical basis, since for single-spin operators, $q=q_i$, according to the generalized Clebsch-Gordan coefficient $\braket{ k_1 q_1 k_2 q_2 \dots k_N q_N | (k_1 k_2 \dots k_N) \{K \} kq }$. By construction, these single-spin operators obey the usual angular momentum commutation relations:
\begin{align}
 [S_\pm,\mathscr{Y}^{(k)q}(\mathbf{S})]=& \sqrt{ k(k+1)-q(q\pm 1)} \mathscr{Y}^{(k)q\pm 1}(\mathbf{S}) \nonumber \\
 [S_z , \mathscr{Y}^{(k)q}(\mathbf{S})] =& q \mathscr{Y}^{(k)q}(\mathbf{S}).
 \label{eq:rot1}
\end{align}
And since these commutation rules hold, as would for any angular momentum operator, any Euler rotation can be applied using multispin tensors of the form $T^{(1)q}_{ \{ K \} }(00 \dots 1_i \dots 00)$, where $1_i$ indicates the component $k_i=1$, and therefore, single-qubit gates can be readily achieved.

The density matrix for $N$ spins can be expanded in the multispin basis:
\begin{equation}
 \rho(t) = \frac{1}{ \prod_{i=1}^N (2S^i+1)} \Bigl[ \sum_{kq\alpha} \phi^k_q(\alpha,t) T^{(k)q}(\alpha) \Bigr]
\label{eq:dens}
\end{equation}
where $\alpha$ is shorthand for all quantum numbers except $k,q$, i.e., the coupling scheme $\{ K \}$ as well as $k_1,\dots, k_N$. In light of (\ref{eq:adj}) and (\ref{eq:orth}) the functions $\phi^k_q(\alpha,t)$ are defined as:
$$ \phi^k_q(\alpha,t) = \mbox{Tr} [ T^{(k)q}(\alpha)^\dagger \rho(t) ] \prod_{i=1}^N (2S^i+1). $$
Each term in the density matrix is of the form (\ref{eq:multis}), which in turn is a summation of products $\mathscr{Y}^{(k_1)q_1}(\mathbf{S}^1) \otimes \mathscr{Y}^{(k_2)q_2}(\mathbf{S}^2) \otimes \dots \otimes  \mathscr{Y}^{(k_N)q_N}(\mathbf{S}^N)$.  A single-qubit gate amounts to applying a unitary constructed from the generator $T^{(1)q}_{ \{ K \} }(00 \dots 1_i \dots 00)\propto \mathbf{S}^i$, which commutes with all tensors $\mathscr{Y}^{(k_j)q_j}(\mathbf{S}^j)$ in the product except $j=i$.  Such rotations of angular momenta $\mathbf{S}^i$ are analogous to the spin 1/2 case.


\subsubsection{Two-Qubit Gates}

In the expression for the density matrix (\ref{eq:dens}) two-qubit gates can affect any term, like the case of single-qubit gates. However, in a product operator $\mathscr{Y}^{(k_1)q_1}(\mathbf{S}^1)  \otimes  \mathscr{Y}^{(k_2)q_2}(\mathbf{S}^2) \otimes \dots \otimes \mathscr{Y}^{(k_N)q_N}(\mathbf{S}^N)$ we need only consider two operators (say 1 and 2), the ones affected by the bilinear coupling Hamiltonian:
\begin{equation}
S_{z}^1 S_{z}^2 = \left\{ \tfrac{ S^1(S^1+1) S^2(S^2+1)}{3} \right\}^{1/2} \mathscr{Y}^{(1)0}(\mathbf{S}^1) \otimes \mathscr{Y}^{(1)0}(\mathbf{S}^2).
\label{eq:s1s2}
\end{equation}
This Hamiltonian will affect $\mathscr{Y}^{(k_1)q_1}(\mathbf{S}^1) \otimes  \mathscr{Y}^{(k_2)q_2}(\mathbf{S}^2)$ in the product operator.
In the multipole formalism, each term in the product has a matrix representation:
\begin{multline}
\mathscr{Y}^{(k_i)q_i}(\mathbf{S}^i) = (i)^{k_i} [ (S^i)(k_i) ]^{1/2} \\
 \times \sum_{M_i M_i'} (-1)^{S^i-M_i} \left(
\begin{matrix}
S^i & k_i & S^i \\
-M_i & q_i & M_i'
\end{matrix}
\right) \ket{ S^i M_i } \bra{ S^i M_i' } 
\label{eq:TmatrixEl}
\end{multline}
where $(S^i) \equiv (2S^i+1)$ and $(k_i)=(2k_i+1)$. For the axial component ($q_i=0$) the selection rule from the Wigner $3j$ symbol implies that $m_i=m_i'$ and the terms are all diagonal in the basis of projection operators $\ket{ S^i M_i } \bra{ S^i M_i' }$.  By extension to the tensor product space, $\mathscr{Y}^{(1)0}(\mathbf{S}^1) \otimes \mathscr{Y}^{(1)0}(\mathbf{S}^2)$ is diagonal in the tensor product basis $\ket{ S^i M_i } \bra{ S^i M_i' }  \otimes \ket{ S^j M_j } \bra{ S^j M_j' }$.  We can therefore exponentiate the operator $\mathscr{Y}^{(1)0}(\mathbf{S}^1) \otimes \mathscr{Y}^{(1)0}(\mathbf{S}^2)$ as follows:
\begin{widetext}
\begin{multline*}
\exp\left( i D t \mathscr{Y}^{(1)0}(\mathbf{S}^1) \otimes \mathscr{Y}^{(1)0}(\mathbf{S}^2) \right) =  \sum_{M_1,M_2} e^{ - i D t [(S^1)(1)(S^2)(1)]^{1/2} (-1)^{S^1-M_1 + S^2 - M_2}  
 \left(
\begin{smallmatrix}
S^1 & 1 & S^1 \\
-M_1 & 0 & M_1
\end{smallmatrix}
\right)  
\left(
\begin{smallmatrix}
S^2 & 1 & S^2 \\
-M_2 & 0 & M_2
\end{smallmatrix}
\right)  } \\
\times \ket{ S^1 M_1 } \bra{ S^1 M_1 }  \otimes \ket{ S^2 M_2 } \bra{ S^2 M_2 } 
\end{multline*}
\end{widetext}
where $D$ is a coupling strength that also absorbs the numerical coefficient of Eq.~(\ref{eq:s1s2}) for convenience. Evolution of a product state such as $\mathscr{Y}^{(k_1)q_1}(\mathbf{S}^1) \otimes \mathscr{Y}^{(k_2)q_2}(\mathbf{S}^2)$ gives:
\begin{widetext}
\begin{multline}
\exp\left( i D t \mathscr{Y}^{(1)0}(\mathbf{S}^1) \otimes \mathscr{Y}^{(1)0}(\mathbf{S}^2) \right) \mathscr{Y}^{(k_1)q_1}(\mathbf{S}^1) \otimes \mathscr{Y}^{(k_2)q_2}(\mathbf{S}^2) \exp\left( - i D t \mathscr{Y}^{(1)0}(\mathbf{S}^1) \otimes \mathscr{Y}^{(1)0}(\mathbf{S}^2) \right)   \\
= \sum_{M_1,M_2} \sum_{M_1',M_2'}  e^{ - i D t [(S^1)(1)(S^2)(1)]^{1/2} (-1)^{S^1-M_1 + S^2 - M_2}  
 \left(
\begin{smallmatrix}
S^1 & 1 & S^1 \\
-M_1 & 0 & M_1
\end{smallmatrix}
\right)  
\left(
\begin{smallmatrix}
S^2 & 1 & S^2 \\
-M_2 & 0 & M_2
\end{smallmatrix}
\right)  } \\
\times e^{  i D t [(S^1)(1)(S^2)(1)]^{1/2} (-1)^{S^1-M_1' + S^2 - M_2'}  
 \left(
\begin{smallmatrix}
S^1 & 1 & S^1 \\
-M_1' & 0 & M_1'
\end{smallmatrix}
\right)  
\left(
\begin{smallmatrix}
S^2 & 1 & S^2 \\
-M_2' & 0 & M_2'
\end{smallmatrix}
\right)  } \\
\times \ket{ S^1 M_1 } \bra{ S^1 M_1 } \mathscr{Y}^{(k_1)q_1}(\mathbf{S}^1) 
 \ket{ S^1 M_1' } \bra{ S^1 M_1' } 
 \otimes \ket{ S^2 M_2 } \bra{ S^2 M_2 }  
\mathscr{Y}^{(k_2)q_2}(\mathbf{S}^2)  
 \ket{ S^2 M_2' } \bra{ S^2 M_2' } 
 \label{eq:rot2}
\end{multline}
\end{widetext}
Expressing the tensors as projectors using Eq.~(\ref{eq:TmatrixEl}) we get:
\begin{widetext}
\begin{multline}
\sigma(t)= (i)^{k_1+k_2} [ (S^1)(k_1) (S^2)(k_2) ]^{1/2} \sum_{M_1,M_2} \sum_{M_1',M_2'}  (-1)^{S^1-M_1 + S^2-M_2}  \\
\times e^{ - i D t [(S^1)(1)(S^2)(1)]^{1/2} (-1)^{S^1-M_1 + S^2 - M_2}  
 \left(
\begin{smallmatrix}
S^1 & 1 & S^1 \\
-M_1 & 0 & M_1
\end{smallmatrix}
\right)  
\left(
\begin{smallmatrix}
S^2 & 1 & S^2 \\
-M_2 & 0 & M_2
\end{smallmatrix}
\right)  } \\
\times  e^{  i D t [(S^1)(1)(S^2)(1)]^{1/2} (-1)^{S^1-M_1' + S^2 - M_2'}  
 \left(
\begin{smallmatrix}
S^1 & 1 & S^1 \\
-M_1' & 0 & M_1'
\end{smallmatrix}
\right)  
\left(
\begin{smallmatrix}
S^2 & 1 & S^2 \\
-M_2' & 0 & M_2'
\end{smallmatrix}
\right)  }  \\
\times \left(
\begin{matrix}
S^1 & k_1 & S^1 \\
-M_1 & q_1 & M_1'
\end{matrix}
\right)  
\left(
\begin{matrix}
S^2 & k_2 & S^2 \\
-M_2 & q_2 & M_2'
\end{matrix}
\right)
\ket{ S^1 M_1 }   \bra{ S^1 M_1' } 
 \otimes \ket{ S^2 M_2 } 
  \bra{ S^2 M_2' }  
 \end{multline}
\end{widetext}
Measurement of a bilinear spin operator $\mathscr{Y}^{(k_3)q_3}(\mathbf{S}^1)  \otimes \mathscr{Y}^{(k_4)q_4}(\mathbf{S}^2)$ is obtained by computing the trace $ \mbox{Tr} \left[ (\mathscr{Y}^{(k_3)q_3}(\mathbf{S}^1) \otimes \mathscr{Y}^{(k_4)q_4}(\mathbf{S}^2))^\dagger \rho(t) \right]$.  Expanding $\mathscr{Y}^{(k_3)q_3}(\mathbf{S}^1)$ and $\mathscr{Y}^{(k_4)q_4}(\mathbf{S}^2)$ in projectors (using Eq.~(\ref{eq:TmatrixEl})) and computing the trace using $\mbox{Tr}(\cdot) = \sum_{ij} \bra{ S^1 M_i }   
 \otimes \bra{ S^2 M_j }  (\cdot) \ket{ S^1 M_i }   
 \otimes \ket{ S^2 M_j }$.  (We have omitted the summation over $S^1$ and $S^2$; all other terms different from $S^1, S^2$ would be zero since $\rho(t)$  only contains terms with $S^1,S^2$.)
\begin{widetext}
\begin{multline}
\phi(t) = [ (S^1)(k_1) (S^2)(k_2) ] \sum_{M_i,M_j} \sum_{M_1,M_2}  (-1)^{S^1-M_1 + S^2-M_2}  e^{ - i D t [(S^1)(1)(S^2)(1)]^{1/2} (-1)^{S^1-M_1 + S^2 - M_2}  
 \left(
\begin{smallmatrix}
S^1 & 1 & S^1 \\
-M_1 & 0 & M_1
\end{smallmatrix}
\right)  
\left(
\begin{smallmatrix}
S^2 & 1 & S^2 \\
-M_2 & 0 & M_2
\end{smallmatrix}
\right)  } \\
\times  e^{  i D t [(S^1)(1)(S^2)(1)]^{1/2} (-1)^{S^1-M_i + S^2 - M_j}  
 \left(
\begin{smallmatrix}
S^1 & 1 & S^1 \\
-M_i & 0 & M_i
\end{smallmatrix}
\right)  
\left(
\begin{smallmatrix}
S^2 & 1 & S^2 \\
-M_j & 0 & M_j
\end{smallmatrix}
\right)  }  \\
\times \left(
\begin{matrix}
S^1 & k_1 & S^1 \\
-M_1 & q_1 & M_i
\end{matrix}
\right)  
\left(
\begin{matrix}
S^2 & k_2 & S^2 \\
-M_2 & q_2 & M_j
\end{matrix}
\right)
\left(
\begin{matrix}
S^1 & k_3 & S^1 \\
-M_1 & q_3 & M_i
\end{matrix}
\right)
\left(
\begin{matrix}
S^2 & k_4 & S^2 \\
-M_2 & q_4 & M_j
\end{matrix}
\right)
 \end{multline}
\end{widetext}
This formula gives the time-evolution for any initial state described by the quantum numbers $S_1,S_2,k_1,q_1,k_2,q_2$.   The special case of a single-spin operator is obtained (for example) by setting $k_2=0,q_2=0$.
Suppose that the initial state is $\mathscr{Y}^{(1)1}(\mathbf{S}^1)$ ($k_1=1$, $q_1=1$, $k_2=0$, $q_2=0$) and we are interested in the amount of $\mathscr{Y}^{(1)1}(\mathbf{S}^1)$ ($k_3=1$, $q_3=1$, $k_4=0$, $q_4=0$).  The product of Wigner $3j$ symbols is nonzero provided that the triangle rules are obeyed (i.e., $0 \le k_1,k_3 \le 2S^1$, $0 \le k_2,k_4 \le 2S^2$).  The Wigner $3j$ symbols lead to the following selection rules: $M_2=M_j$ and $M_1=M_i+1$, causing the sums over $M_i$ and $M_j$ to collapse. Inside the argument of the exponential, we may simplify the term:
\begin{multline}
\left\{
   \left(
\begin{smallmatrix}
S^1 & 1 & S^1 \\
-M_1 & 0 & M_1
\end{smallmatrix}
\right)  
 - \left(
\begin{smallmatrix}
S^1 & 1 & S^1 \\
1-M_1 & 0 & M_1-1
\end{smallmatrix}
\right)   \right\} 
\left(
\begin{smallmatrix}
S^2 & 1 & S^2 \\
-M_2 & 0 & M_2
\end{smallmatrix}
\right) \\
= \frac{ (-1)^{M_1+S^1 + M_2 + S^2} M_2 }{ \sqrt{ [S^1(S^1+1)(2S^1+1)] [S^2(S^2+1)(2S^2+1)] } }
 \end{multline}
 provided that $-S^1 \le M_1-1 \le S^1$ and $-S^2 \le M_2 \le S^2$. We get:
\begin{widetext}
\begin{multline}
\mathcal{C}(t)  = [ (S^1)(1) (S^2)(0) ] \sum_{M_1,M_2}  (-1)^{S^1-M_1 + S^2-M_2}  \\
 \times  e^{ - i D t [(S^1)(1)(S^2)(1)]^{1/2} 
(-1)^{M_1+M_2+S^1+S^2} M_2  [S^1(S^1+1)(2S^1+1) S^2(S^2+1)(2S^2+1)]^{-1/2}   } \\ 
\times \frac{ (M_1-1-S^1) (S^1-M_1)! }{ 2 S^1(1+3S^1+2[S^1]^2)(1+2S^2) (M_1-1+S^1)!} 
\label{eq:harm1}
 \end{multline}
\end{widetext}
where $[S^1]^2$ denotes the square of $S^1$ (spin 1 magnitude).  For fixed $S^1, S^2$ this equation describes a sum of oscillatory functions, each with a slightly different frequency ($M_1, M_2$ dependence) and amplitudes.  The initial coherence ($q_1=1$) returns to its original state periodically, at a frequency that can be read out from the argument of the exponential; the factor $M_2$ determines the particular harmonic for each term in the summation.


The limit of large spin is obtained by taking $S^1=S^2=S$ and letting $S \rightarrow \infty$ (i.e., $2S+1 \approx S$, etc) and recalling that $D$ is proportional to $[S]^2$ (see Eq.~\ref{eq:s1s2}).   The argument of the exponential (evolution frequency) is then seen to be proportional to $([S]^2 \cdot S / [S]^3)M_2 = M_2$ and the frequencies  are bounded by $|M_2|\le S$.

On the other hand, suppose that we are interested in the amount of $\mathscr{Y}^{(1)1}(\mathbf{S}^1)  \otimes \mathscr{Y}^{(1)0}(\mathbf{S}^2)$ ($k_3=1$, $q_3=1$, $k_4=1$, $q_4=0$).  
The Wigner $3j$ symbols again lead to the following selection rules: $M_2=M_j$ and $M_1=M_i+1$, causing the sums over $M_i$ and $M_j$ to collapse.  We get:
\begin{multline}
\mathcal{S}(t) = [ (S^1)(1) (S^2)(0) ] \sum_{M_1,M_2}  (-1)^{S^1-M_1 + S^2-M_2}   \\
 \times e^{ - i D t [(S^1)(1)(S^2)(1)]^{1/2} 
(-1)^{M_1+M_2+S^1+S^2} M_2 } \\
 \qquad {}^{ \times [S^1(S^1+1)(2S^1+1) S_2(S^2+1)(2S^2+1)]^{-1/2}  }  \\  
 \cdot \frac{ \times M_2 (1-M_1+S^1) (S^1+M_1)! }{ 2 S_1(1+3S^1+2[S^1]^2) \sqrt{S^2(1+S^2)} (1+2S^2) (M_1-1+S^1)!}  \\
\label{eq:osc1}
\end{multline}
The harmonic content is the same as in Eq.~(\ref{eq:harm1})  whereas the amplitudes are slightly different. This describes an evolution of the form:
\begin{multline}
 \mathscr{Y}^{(1)1}(\mathbf{S}^1) \xlongrightarrow[]{D \mathscr{Y}^{(1)0}(\mathbf{S}^1) \otimes \mathscr{Y}^{(1)0}(\mathbf{S}^2)}  \\
  \mathscr{Y}^{(1)1}(\mathbf{S}^1) \mathcal{C}(t)  + \mathscr{Y}^{(1)1}(\mathbf{S}^1) \otimes \mathscr{Y}^{(1)0}(\mathbf{S}^2) \mathcal{S}(t)
\end{multline}
where $\mathcal{C}(t) $ and $\mathcal{S}(t)$ are periodic oscillatory functions whose frequency content consists of harmonics of the fundamental frequency, which is proportional to $D$ (see Eqs.~\ref{eq:harm1} and~\ref{eq:osc1}).   This rule is analogous to the rule for $J$-coupling evolution in NMR, 
$$ I_{ \pm }^k \rightarrow I_{\pm}^k \cos \pi J_{kl} \tau \mp i 2 I_{\pm}^k I_{z}^l \sin \pi J_{kl} \tau $$
except that the periodic coefficients contain harmonics because of the multipoles.  Similarly, it can be shown that entangled states like Bell states can be created analogously to the case of spin-1/2 systems.   The important point to realize is that while quantum information can ``leak'' into other tensor components (not all combinations of possible initial and final states are shown here), the periodic nature of the coefficients shows that 2-qubit gates can be realized even amongst large spins.

\subsection{Definition of Gates and Computational Basis for Molecular Magnets\label{appC}}

In this section we define the computational basis and sequences of operations that lead to a universal set of gates.  The Liouville space of operators for arbitrary spins $S$ is spanned by the basis of irreducible tensor operators $\mathscr{Y}^{(k)q}(\mathbf{S})$, $0 \le k \le 2S$.  The space is also spanned by the outer products $\ket{SM}\bra{SM'}$.  The relationship between $\ket{SM}\bra{SM'}$ and irreducible tensors $\mathscr{Y}^{(k)q}(\mathbf{S})$ is:
\begin{multline}
\ket{SM}\bra{SM'} = (-1)^{S-M} (2S+1)^{-1/2}\\
 \times \sum_{k=0}^{2S} \sum_{q=-k}^k (-i)^k (2k+1)^{1/2} \left(
\begin{matrix}
S & k & S \\
-M & q & M'
\end{matrix}
\right) \mathscr{Y}^{(k)q}(\mathbf{S}).
\label{eq:expp} 
\end{multline}
We define the computational basis for arbitrary spins in terms of the qubit states shown in Table~\ref{tab:compbasis}.  The transformation rules of multipole tensors are listed in Table~\ref{tab:action} whereas a possible implementation of quantum gates is show in Table~\ref{tab:gates}.  In these tables the operator ket notation $\dket{B}$ denotes the operator itself ($B$), whereas the bra $\dbra{A}$ denotes the adjoint $A^\dagger$ and $\dbra{A}\dket{B}$ forms an inner product, often taken to be $\mbox{Tr}[A^\dagger B]$.
\begin{table}[h!]
 \begin{tabular}{|| m{5em} | m{18em} ||}
 \hline
 Single-Spin State   & Multipole Representation  \\ [0.5ex]
 \hline\hline
 $\dket{0}$, $\dket{z}$ & $ \sum\limits_{k=0}^{2S} \phi_0^k
\mathscr{Y}^{(k)0}(\mathbf{S}) = \sum\limits_{k=0}^{2S} \phi_0^k
\dket{k0}  $  \\
 \hline
 $\dket{1}$, $\dket{-z}$&   $\sum\limits_{k=0}^{2S} (-1)^k \phi_0^k
\mathscr{Y}^{(k)0}(\mathbf{S})  $  \\
 \hline
 $\dket{x}$&  $\sum\limits_{k=0}^{2S} 
\phi^k_0 \sum\limits_{q=-k}^k \mathscr{D}^{(k)}_{q0}(0,-\tfrac{\pi}{2},0) \mathscr{Y}^{(k)q}(\mathbf{S})  $  \\
 \hline
  $\dket{y}$&  $\sum\limits_{k=0}^{2S} 
\phi^k_0 \sum\limits_{q=-k}^k \mathscr{D}^{(k)}_{q0}(\tfrac{\pi}{2},\tfrac{\pi}{2},0) \mathscr{Y}^{(k)q}(\mathbf{S})  $  \\ [1ex]
 \hline
\end{tabular}
\caption{Computational basis for arbitrary spins in terms of multipoles. The numerical coefficients $\phi_0^k$ defining the fully polarized state are obtained from Eq.~(\ref{eq:expp}) by setting $M,M'=S$.  Here, $\mathscr{D}^{(k)}_{qq'}(\Omega)$ are Wigner $D$-matrices.   Alternative notation for $\mathscr{Y}^{(k)q}(\mathbf{S}) $ is $\dket{kq}$. \label{tab:compbasis} }
\end{table}

\begin{table}[h!]
\begin{tabular}{|| m{3em}| m{12em} | m{12em} ||}
 \hline
 Action & Operator & Transformation Rule  \\ [0.5ex]
 \hline\hline
   $ R_z^A(\phi) $ & $e^{\frac{-i}{2} \phi \mathscr{Y}^{(1)0}(\mathbf{S}^A)}$
 & $\dket{x} \rightarrow \dket{x} \widetilde{\cos}{\phi}-i \dket{y} \widetilde{\sin}{\phi}  $  \\
 \hline
 $ R_x^A(\phi)$ & $ e^{\frac{-i}{2} \phi
(\mathscr{Y}^{(1)1}(\mathbf{S}^A)-\mathscr{Y}^{(1)-1}(\mathbf{S}^A))} $
 & $\dket{y} \rightarrow \dket{y}\widetilde{\cos}{\phi}-i \dket{z}\widetilde{\sin}{\phi}  $   \\
 \hline
  $R_y^A(\phi)$ & $ e^{\frac{-i}{2} \phi
(\mathscr{Y}^{(1)1}(\mathbf{S}^A)+\mathscr{Y}^{(1)-1}(\mathbf{S}^A))} $
 & $\dket{x} \rightarrow \dket{x} \widetilde{\cos} {\phi}+i \dket{z} \widetilde{\sin} {\phi}  $  \\
 \hline
 $U_{zz}(\frac{\pi}{2})$ & $ e^{\frac{-i \pi}{4}
\mathscr{Y}^{(1)0} (\mathbf{S}^A) \scriptstyle{\otimes} \mathscr{Y}^{(1)0} (\mathbf{S}^B)}$
&  $ \dket{x}^{A} \otimes \mathds{1}^{B}\rightarrow \dket{y}^{A}
\otimes \dket{z}^{B}   $ \\ [1ex]
 \hline
\end{tabular}
\caption{ \label{tab:action} Transformation rules for multipole tensors representing the computational basis states.  The notation $\widetilde{\cos}$ and $\widetilde{\sin}$ is a reminder that individual tensor components are evolved according to the  harmonics generated by the transformation rules of tensor operators.  For single-qubit gates, this function is obtained from the commutation relations~(\ref{eq:rot2}).  For two-qubit gates the periodic functions can be read off from the argument to the exponential function in Eq.~(\ref{eq:osc1}) and (\ref{eq:harm1}). }
\end{table}

\begin{table}[h!]
\begin{tabular}{|| m{5em}| m{5em} | m{12em} ||}
 \hline
 Gate & Symbol & Unitary operator  \\ [0.5ex]
 \hline\hline
 X bit-flip & \Qcircuit @C=1em @R=.7em { & \gate{X} & \qw} &
  $R_x(\pi)$  \\
 \hline
 Z phase-flip & \Qcircuit @C=1em @R=0.7em { & \gate{Z} & \qw}
 & $R_z(\pi)$  \\
 \hline
Hadamard & \Qcircuit @C=1em @R=.7em { & \gate{H} & \qw}
&  $R_z(\frac{\pi}{2})\cdot R_x(\frac{\pi}{2})\cdot R_z(\frac{\pi}{2})   $  \\
 \hline
 CNOT & \Qcircuit @C=1em @R=.7em {
& \ctrl{1} & \qw \\ & \targ  &\qw
}&  $\sqrt{i} R^A_z(\frac{\pi}{2}) \cdot R^B_z(\frac{-\pi}{2}) \cdot
R^B_x(\frac{\pi}{2}) \cdot U_{zz}(\frac{1}{2\| \mathtensor{D}_{AB}\|}) \cdot
R^B_y(\frac{\pi}{2})$ \\ [1ex]
 \hline
\end{tabular}
\caption{ \label{tab:gates} Implementation of the elementary gates as sequences of selective rotations.  Each operator except $U$ acts on the subspace of a single qudit. }
\end{table}

\subsection{Extended Survival Times in Non-Interacting Spins\label{appA}}

Entangled spins within a localized ensemble decay faster than uncoupled spins, whereas uncoupled spin ensembles survive longer than individual spins.  Indeed, non-entangled spins decay independently and the quantum information stored within the polarization of the ensemble remains encoded over longer periods.  Here we give a rough estimate of the lifetime enhancement provided by an ensemble relative to individual spins based on a probabilistic argument. Let $X$ be a random variable describing the survival time of the spin. A single spin decays at a rate $\lambda$. Assuming a Poisson process with distribution function of the arrival times  $\mathbb{P}(X>x)=e^{-\lambda x}$. The corresponding PDF is $\frac{d\mathbb{P}(X\le x)}{dx}  = \lambda e^{-\lambda x}$. Thus, the average arrival time (lifetime) for this single spin is $\mathbb{E}(X) = \int_0^\infty x \lambda e^{-\lambda x} dx = \lambda^{-1}$, as expected. Once this spin relaxes, the quantum state is lost; this is the main disadvantage of using single spins for quantum computing. When the state has $n$ spins, each of which decays in the same way (but independently), the probability of the quantum state getting lost can be obtained by considering the random variable $Y = \max \left\{ X_1,\dots,X_n \right\}$, where $X_i$ are {\it iid} describing the survival time of each of the $n$ spins. The lifetime of the $n$-spins state is determined by relaxation of the last spin, hence our interest in the maximum of $X_1,\dots,X_n$. Assume that the spins are non-interacting, the joint probability distribution is $\mathbb{P}( Y \le x) = \mathbb{P}(X_1 \le x, \dots, X_n \le x) = \prod_{i=1}^n \mathbb{P}(X_i \le x)$, where each $\mathbb{P}(X_i \le x)=1-\mathbb{P}(X_i > x)=1-e^{-\lambda x}$ and the product form assumes the spins are independent (non-interacting).  The lifetime of the uncoupled $n$-spins system is:
$$ \mathbb{E}(Y) = \int_0^\infty y  \mathbb{P}(Y \in dy)  = \int_0^\infty y n (1-e^{-\lambda y})^{n-1} \lambda e^{-\lambda y} dy, $$
which approaches $n \lambda^{-1}$ as $n$ increases. Thus, the spin-coherent state lifetime is $n$ times longer than in the case of a single spin. This is an important advantage over single spins or entangled spins in an ensemble. We note that the lifetime of the ``quantum memory'' does not equal $n \lambda^{-1}$  since its usefulness rapidly disappears below the threshold of spin noise ($\propto \sqrt{n}$).

\subsection{Dissipation in Many-Body Coupled Spin Systems\label{appA2}}


We now discuss the dissipation rates ($\lambda$ in the previous section).  The E-qubit requires weak couplings among the spins of the ensemble.   Although challenging, residual couplings should be removed.   Nuclear spins can be decoupled efficiently using a combination of sample spinning and rotor-synchronized RF pulses.  Narrowing of lines down to the sub-hertz range has been demonstrated~\cite{Boutis03,levitt2007symmetry}.  For electrons, however, sample spinning is not an option, whereas pulsed decoupling has limited value.  Material engineering may be the best option.  Sellars et al.~\cite{ranvcic2018coherence}  demonstrated that co-doping with oxygen can reduce line widths down to 10 Hz. If the spins are decoupled, the lifetime of a quantum state encoded in the E-qubit will  be extended relative to that of a single spin.  

In the presence of residual pairwise couplings a one-body state coherently evolves into many-body states.  The latter decay much faster than single-body states. Pulsed decoupling can help prevent this coherent evolution. In this section we show the detrimental effects of decoherence on many-body states.   The conclusion we draw is one that strongly advises against allowing the formation of correlated many-body states, either intentionally via quantum state preparation, or unintentionally via residual spin-spin couplings.

Consider a general Hamiltonian of the form 
$ \mathcal{H} = \mathcal{H}_Z + \mathcal{H}_{SB} + \mathcal{H}_B$,
where $\mathcal{H}_Z$ is the unperturbed spin part (e.g. Zeeman interaction), $\mathcal{H}_{SB}$ is the spin-bath interaction including any spin-spin interactions and $\mathcal{H}_B$ is the bath Hamiltonian.  We denote the corresponding Liouvillian superoperator as $\mathscr{L}=\mathscr{L}_Z + \mathscr{L}_{SB} + \mathscr{L}_B$.
Consider a set of operators $\{ F_k\}_{k=1}^m$ and an inner product $(A,B)=\mbox{Tr}(A^\dagger B)$, often denoted $\langle\langle A|B \rangle\rangle$. The operators $F_k$ are orthogonal, $\langle\langle F_i|F_j \rangle\rangle=0$ for $i \ne j$.  A projection operator is defined as $\pi = \sum_{k=1}^m \frac{ | F_k \rangle\rangle  \langle\langle F_k | }{ \langle\langle F_k | F_k \rangle\rangle }$. It is customary to transform the Liouvillian to the interaction representation generated by $\mathscr{L}_Z$ and write the resulting Liouvillian as $\mathscr{L}^*(t) \equiv \mathscr{L}_B + \mathscr{L}_{SB}(t)$, where $*$ denotes the interaction representation and $\mathscr{L}_{SB}(t) \equiv e^{i t \mathscr{L}_Z} \mathscr{L}_{SB}$. To study irreversible processes, projection operator methods yield a quantum master equation for an observable $F_k$:
\begin{widetext}
\begin{align*}
\frac{d}{dt} \langle\langle F_k|\rho(t) \rangle\rangle =& -i \langle\langle F_k | \mathscr{L}^*(t) \pi | \rho(t) \rangle\rangle - i \langle\langle F_k| \mathscr{L}^*(t) Te^{-i\int_{0}^t d\tau (1-\pi) \mathscr{L}^*(\tau) }
 (1-\pi) |\rho(0) \rangle\rangle \\
 & - \sum_{j=1}^m \int_0^t dt' \frac{  \langle\langle F_k| \mathscr{L}^*(t) Te^{-i\int_{t'}^t d\tau (1-\pi) \mathscr{L}^*(\tau) }
 (1-\pi) \mathscr{L}^*(t') | F_j \rangle\rangle }{ \langle\langle F_j | F_j \rangle\rangle} \langle\langle F_j|\rho(t') \rangle\rangle
\end{align*}
\end{widetext}
where $T$ denotes Dyson time-ordering of the exponential.
Insertion of the expression for the projection operator $\pi$ into the first term shows that it causes coherent evolution (Bloch term). The second term can be made to vanish with a proper choice of initial conditions.  The last term is the dissipative term we are interested in. Under some assumptions (weak collision limit, short correlation time limit, average Liouvillian, stationarity of the memory kernel), $Te^{-i\int_{t'}^t d\tau (1-\pi) \mathscr{L}^*(\tau) }$ is approximated by $e^{-i(t-t')(1-\pi)\mathscr{L}^* }$, where $\mathscr{L}^*$ is a time-averaged Liouvillian in the interaction representation.  It is customary to drop the Zeeman and spin-bath parts ($\| \mathscr{L}_B \| \gg \|\mathscr{L}_Z + \mathscr{L}_{SB}\|$), leaving only the bath part, i.e. $\mathscr{L}^* \approx \mathscr{L}_B$, in the exponential. Since there is no spin part left, the projector term is dropped leaving $e^{-i(t-t')\mathscr{L}_B}$.  Finally, the term $(1-\pi) \mathscr{L}^*(t') | F_j \rangle\rangle$ gives two terms: $\mathscr{L}^*(t') | F_j \rangle\rangle$ and $-\pi \mathscr{L}^*(t') | F_j \rangle\rangle$.  The second term is not of interest in the derivation of dissipation rates because it describes a product of average frequencies; we are instead interested in deviations from  averages. We therefore consider the first term.  The dissipation term in this wide-sense stationary Redfield limit reads:
\begin{align*}
 & \sum_{j=1}^m \int_0^\infty d\tau \frac{ \langle\langle F_k| \mathscr{L}^*(0) e^{i\tau \mathscr{L}_B}
\mathscr{L}^*(\tau) | F_j \rangle\rangle }{ \langle\langle F_j | F_j \rangle\rangle } \langle\langle F_j|\rho(t) \rangle\rangle \\
 & \qquad \equiv \sum_{j=1}^m \langle\langle F_j|\rho(t) \rangle\rangle \int_0^\infty \mathcal{K}_{kj}(\tau) d\tau,
 \end{align*}
where $W(F_k) \equiv  \int_0^\infty \mathcal{K}_{kj}(\tau) d\tau$ is a dissipation rate for the state $F_k$. The spin-bath interaction is taken to be a sum of pairwise couplings.
In the interaction representation, it acquires a phase factor $e^{- i \omega q \tau}$:
\begin{align*}
 \mathcal{H}_{SB}^*(\tau) =& \sum_{i<j} \sum_{k=0,1,2} \sum_{q=-2}^2 (-1)^q A_{kq}( \mathbf{S}^i, \mathbf{S}^j) T_{k,-q}(\mathbf{S}^i,\mathbf{S}^j) \\
& \qquad  \times e^{- i \omega q \tau}
\end{align*}
where we used the shorthand notation
$ T_{k,-q}(\mathbf{S}^i,\mathbf{S}^j)$ for the multispin tensor $T_{k,-q}(000 \underbrace{\dots 1\dots }_{i\mathrm{th~pos.}} \underbrace{\dots 1 \dots }_{j\mathrm{th~pos.}} 000)$, which we will denote as $T^{kq}(k_1k_2)$ with $k_1=k_2=1$.

\subsubsection{One-Body States}

The one-body state describes a spin among an ensemble of uncoupled spins (E-qubit), or weakly coupled spins undergoing decoupling.   It also describes an initial state ($t=0$) where the spins did not undergo coherent evolution. It is described by a vector (rank 1) tensor operator corresponding to a spin of interest (denoted  $m$ henceforth):
$$ |F_m \rangle\rangle = \mathscr{Y}^{(1)q}(\mathbf{S}^m) $$
where $q=-1,0,1$. Spin-lattice  ($T_1$) relaxation corresponds to the case $q=0$ whereas spin-spin relaxation corresponds to $q=\pm 1$. We now compute the numerator of the memory function, 
$\langle\langle F_k| \mathscr{L}^*(0) e^{i\tau \mathscr{L}_B}
\mathscr{L}^*(\tau) | F_j \rangle\rangle$.
We start with the commutator:
\begin{align*}
\mathscr{L}^*(\tau) F_m & \equiv [\mathcal{H}_{SB}^*(\tau),F_m ] \\
& = \sum_{i<j} \sum_{k=0,1,2} \sum_{q=-2}^2 (-1)^q A_{kq}( \mathbf{S}^i, \mathbf{S}^j) \\
& \quad \times [ T_{2,-q}(\mathbf{S}^i,\mathbf{S}^j) , \mathscr{Y}^{(1)q}(\mathbf{S}^m)  ]  e^{- i \omega q \tau}.
\end{align*}
Substituting
\begin{align*}
T^{kq}(k_1k_2) &= [(2S^1+1)(2S^2+1)]^{-1/2} \\
& \times \sum_{q_1 q_2} (-1)^{k_1 - k_2 + k} \sqrt{(2k+1)}  \\
 & \times (-1)^{k-q} \left(
\begin{matrix}
k & k_1 & k_2 \\
-q & q_1 & q_2
\end{matrix} \right)
\mathscr{Y}^{(k_1)q_1} (\mathbf{S}^1) 
\mathscr{Y}^{(k_2)q_2} (\mathbf{S}^2) 
\end{align*}
we get:
$$ \mathscr{L}^*(\tau) F_m  = \frac{1}{2} \sum_{i,j} \sum_{k=0,1,2} \sum_{q=-2}^2 (-1)^q A_{kq}( \mathbf{S}^i, \mathbf{S}^j) $$
$$ \times \sum_{q_1 q_2} (-1)^{k_1 - k_2 + k} \sqrt{(2k+1)} [(2S^i+1)(2S^j+1)]^{-1/2}  
(-1)^{k-q} $$
$$ \times \left(
\begin{matrix}
k & k_1 & k_2 \\
-q & q_1 & q_2
\end{matrix} \right)
[ \mathscr{Y}^{(k_1)q_1} (\mathbf{S}^i) 
\mathscr{Y}^{(k_2)q_2} (\mathbf{S}^j)  , \mathscr{Y}^{(1)q}(\mathbf{S}^m)  ]  e^{- i \omega q \tau}.  $$
Substitution of the commutator
\begin{align*}
 [\mathscr{Y}^{(l)m}(\mathbf{S}), & \mathscr{Y}^{(k)q}(\mathbf{S})] = 2 \sum_{k'q'} \frac{ \phi(klk') }{ (2S+1) } \\
  & \times \langle \langle \mathscr{Y}^{(k')q'}(\mathbf{S}) | \mathscr{Y}^{(l)m}(\mathbf{S}) | \mathscr{Y}^{(k)q}(\mathbf{S}) \rangle \rangle \mathscr{Y}^{(k')q'}(\mathbf{S}) 
\end{align*}
where $\phi(klm')$ is $1$ if $k+k'+l$ is odd and zero otherwise,
\begin{align*}
\langle \langle \mathscr{Y}^{(k')q'}(\mathbf{S}) & | \mathscr{Y}^{(l)m}(\mathbf{S}) | \mathscr{Y}^{(k)q}(\mathbf{S}) \rangle \rangle  \\
 = & \mbox{Tr}  \left\{ \mathscr{Y}^{(k')q'}(\mathbf{S})^\dagger \mathscr{Y}^{(l)m}(\mathbf{S})  \mathscr{Y}^{(k)q}(\mathbf{S})  \right\} \\
=& (-1)^{k'-q'} \left( \begin{matrix} k' & l & k \\
-q' & m & q 
\end{matrix}
\right) \langle \langle k' \| \mathscr{Y}^{(l)} \| k \rangle \rangle
\end{align*}
(Wigner-Eckart theorem) and 
\begin{align*}
 \langle \langle & k' \| \mathscr{Y}^{(l)} \| k \rangle \rangle = (-1)^{l+k+k'+2S} (i)^{k+k'+l} \\
 & \times [(2l+1)(2k+1)(2S+1)(2k'+1)]^{1/2} 
\left\{ \begin{matrix} k' & l & k \\
S & S & S 
\end{matrix}
\right\},
\end{align*}
the reduced matrix element, in terms of Wigner $6-j$ symbols, yields:
\begin{align*}
 & \mathscr{L}^*(\tau) F_m  = \sum_{i<j} \sum_{k=0,1,2} \sum_{q=-2}^2 (-1)^q A_{kq}( \mathbf{S}^i, \mathbf{S}^j) (-1)^{k-q} \\
 & \times \sum_{q_1 q_2} (-1)^{k_1 - k_2 + k} \sqrt{(2k+1)} [(2S^i+1)(2S^j+1)]^{-1/2}  \\
& \times \left(
\begin{matrix}
k & k_1 & k_2 \\
-q & q_1 & q_2
\end{matrix} \right)
[ \mathscr{Y}^{(k_1)q_1} (\mathbf{S}^i) 
\mathscr{Y}^{(k_2)q_2} (\mathbf{S}^j)  , \mathscr{Y}^{(1)q}(\mathbf{S}^m)  ]e^{- i \omega q \tau}.
\end{align*}
Acting on $\mathscr{L}^*(\tau)F$ with the bath propagator $e^{ i\mathscr{L}_B \tau}$ introduces the time dependence $A_{kq}( \mathbf{S}^i, \mathbf{S}^j)  \rightarrow A_{kq}( \mathbf{S}^i, \mathbf{S}^j)(\tau)$. Now,
\begin{align*}
[ & \mathscr{Y}^{(k_1)q_1} (\mathbf{S}^i)   
\mathscr{Y}^{(k_2)q_2} (\mathbf{S}^j),\mathscr{Y}^{(1)q}(\mathbf{S}^m)] =
\frac{2}{ (2S^m+1) } \sum_{k'q'}  \\
& \times \Bigl\{ \delta_{im}  
 \phi(1k_1 k')  \langle \langle \mathscr{Y}^{(k')q'}(\mathbf{S}^m) | \mathscr{Y}^{(k_1)q_1}(\mathbf{S}^m) | \mathscr{Y}^{(1)q}(\mathbf{S}^m) \rangle \rangle \\
 & \times \mathscr{Y}^{(k')q'}(\mathbf{S}^m) \mathscr{Y}^{(k_2)q_2}(\mathbf{S}^j) \\
& + \delta_{jm} 
 \phi(1k_2 k')  \langle \langle \mathscr{Y}^{(k')q'}(\mathbf{S}^m) | \mathscr{Y}^{(k_2)q_2}(\mathbf{S}^m) | \mathscr{Y}^{(1)q}(\mathbf{S}^m) \rangle \rangle \\
 & \times \mathscr{Y}^{(k')q'}(\mathbf{S}^m)  \mathscr{Y}^{(k_1)q_1}(\mathbf{S}^i) \Bigr\}, 
\end{align*}
where $i \ne j$ (LHS) implying that $j \ne m$ in the first term and $i \ne m$ in the second term (RHS). The final expression is:
\begin{align}
& e^{\mathscr{L}_B \tau } \mathscr{L}^*(\tau) F_m = \sum_{i<j} \sum_{k=0,1,2} \sum_{q=-2}^2 (-1)^q A_{kq}( \mathbf{S}^i, \mathbf{S}^j)(\tau) \nonumber \\
& 
\sum_{q_1 q_2} (-1)^{k_1 - k_2 + k} \sqrt{(2k+1)} [(2S^i+1)(2S^j+1)]^{-1/2}  
(-1)^{k-q}  \nonumber \\
 & \times \left(
\begin{matrix}
k & k_1 & k_2 \\
-q & q_1 & q_2
\end{matrix} \right) \frac{2 e^{-i\omega q \tau}}{ (2S^m+1) } \sum_{k'q'} \Bigl\{ \delta_{im}  
 \phi(1k_1 k')  \nonumber \\
 & \times  \langle \langle \mathscr{Y}^{(k')q'}(\mathbf{S}^m) | \mathscr{Y}^{(k_1)q_1}(\mathbf{S}^m) | \mathscr{Y}^{(1)q}(\mathbf{S}^m) \rangle \rangle  \nonumber \\
 & \times \mathscr{Y}^{(k')q'}(\mathbf{S}^m)  \mathscr{Y}^{(k_2)q_2}(\mathbf{S}^j)  \nonumber \\
& + \delta_{jm} 
 \phi(1k_2 k')  \langle \langle \mathscr{Y}^{(k')q'}(\mathbf{S}^m) | \mathscr{Y}^{(k_2)q_2}(\mathbf{S}^m) | \mathscr{Y}^{(1)q}(\mathbf{S}^m) \rangle \rangle  \nonumber \\
 & \times \mathscr{Y}^{(k')q'}(\mathbf{S}^m)  \mathscr{Y}^{(k_1)q_1}(\mathbf{S}^i) \Bigr\}
  \label{eq:LR}
\end{align}
where $k_1=k_2=1$.
We also compute the quantity $(\mathscr{L}^*(0) F_{\tilde{m}} )^\dagger$ by conjugation using $\mathscr{Y}^{(k)q}{}^\dagger(\mathbf{S}) = (-1)^{k-q}\mathscr{Y}^{(k)-q}(\mathbf{S})$.  Thus,
\begin{align}
& (\mathscr{L}^*(0)F_m )^\dagger =  \sum_{ \tilde{i} < \tilde{j} } \sum_{ \tilde{k}=0,1,2 } \sum_{ \tilde{q}=-2 }^2 (-1)^{\tilde{q}} A_{ \tilde{k} \tilde{q} }( \mathbf{S}^{\tilde{i}}, \mathbf{S}^{\tilde{j}} )(0)^\dagger  \nonumber  \\
& \times \sum_{ \tilde{q}_1 \tilde{q}_2} (-1)^{k_1 - k_2 + \tilde{k} } \sqrt{(2 \tilde{k}+1)} [(2S^{ \tilde{i} }+1)(2S^{ \tilde{j} }+1)]^{-1/2}    \nonumber \\
 & \times (-1)^{ \tilde{k}-\tilde{q} }  \left(
\begin{matrix}
\tilde{k} & k_1 & k_2 \\
-\tilde{q} & \tilde{q}_1 & \tilde{q}_2
\end{matrix} \right) 
\frac{2}{ (2S^{\tilde{m}}+1) } \sum_{ \tilde{k}' \tilde{q}' } \Bigl\{ \delta_{ \tilde{i} \tilde{m} }  
 \phi(1k_1 \tilde{k}')  \nonumber  \\
 & \times \langle \langle \mathscr{Y}^{ (\tilde{k}') \tilde{q}'}(\mathbf{S}^{\tilde{m}}) | \mathscr{Y}^{(k_1) \tilde{q}_1}(\mathbf{S}^{\tilde{m}}) | \mathscr{Y}^{(1)\tilde{q}}(\mathbf{S}^{\tilde{m}}) \rangle \rangle  \nonumber  \\
 & \times \mathscr{Y}^{(\tilde{k}')\tilde{q}'}(\mathbf{S}^{\tilde{m}})^\dagger  \mathscr{Y}^{(k_2) \tilde{q}_2}(\mathbf{S}^{\tilde{j}})^\dagger  \nonumber \\
& + \delta_{\tilde{j}\tilde{m}} 
 \phi(1k_2 \tilde{k}') \langle \langle \mathscr{Y}^{(\tilde{k}')\tilde{q}'}(\mathbf{S}^{\tilde{m}}) | \mathscr{Y}^{(k_2) \tilde{q}_2}(\mathbf{S}^{\tilde{m}}) | \mathscr{Y}^{(1)\tilde{q}}(\mathbf{S}^{\tilde{m}}) \rangle \rangle  \nonumber  \\
 & \times \mathscr{Y}^{(\tilde{k}')\tilde{q}'}(\mathbf{S}^{\tilde{m}})^\dagger \mathscr{Y}^{(k_1) \tilde{q}_1}(\mathbf{S}^{\tilde{i}})^\dagger \Bigr\}, 
 \label{eq:LL}
\end{align}
where $k_1=k_2=1$.  We have relabeled the index $m$ as $\tilde{m}$ to allow for the possibility of cross-relaxation. 
Next, we compute the inner product:
\begin{align}
\bigl( \mathscr{L}^*(0) F_{\tilde{m}} , & e^{ i\mathscr{L}_B \tau } \mathscr{L}^*(\tau) F_m \bigr) =   \nonumber \\ 
& \mbox{Tr} \bigl[ (\mathscr{L}^*(0) F_{\tilde{m}})^\dagger e^{i\mathscr{L}_B \tau} \mathscr{L}^*(\tau) F_m \bigr].
\label{eq:IP}
\end{align}
To this end, we use the trace formula:
$$ \mbox{ Tr} \{ \mathscr{Y}^{(k)q}(\mathbf{S})^\dagger \mathscr{Y}^{(k')q'}(\mathbf{S}) \}  = (2S+1) \delta_{kk'} \delta_{qq'}. $$
Upon multiplying $(\mathscr{L}^*(0) F_{\tilde{m}})^\dagger$ and $e^{i\mathscr{L}_B \tau} \mathscr{L} ^*(\tau) F_m$ we get four terms.
The operator part gives:
\begin{align}
 & \mbox{Tr} \left[ \mathscr{Y}^{(\tilde{k}')\tilde{q}'}(\mathbf{S}^{\tilde{m}})^\dagger  \mathscr{Y}^{(k_2)\tilde{q}_2}(\mathbf{S}^{\tilde{j}})^\dagger 
\mathscr{Y}^{(k')q'}(\mathbf{S}^m)  \mathscr{Y}^{(k_2)q_2}(\mathbf{S}^j)
\right]   \nonumber \\
 & = (2S^m+1)(2S^j+1) \bigl[ \delta_{ \tilde{m} m} \delta_{ \tilde{j} j}  \delta_{ \tilde{k}' k'} \delta_{ \tilde{q}' q'} \delta_{\tilde{q}_2 q_2}   \nonumber \\
 & \quad + \delta_{ \tilde{j} m } \delta_{ \tilde{m} j} \delta_{ 1 k'} \delta_{ \tilde{q}_2 q'}  \delta_{\tilde{k}' 1}  \delta_{ \tilde{q}' q_2} \bigr]
 \label{eq:trace1}
 \end{align}
\begin{align}
& \mbox{Tr} \left[ \mathscr{Y}^{(\tilde{k}')\tilde{q}'}(\mathbf{S}^{\tilde{m}})^\dagger  \mathscr{Y}^{(k_2)\tilde{q}_2}(\mathbf{S}^{\tilde{j}})^\dagger 
\mathscr{Y}^{(k')q'}(\mathbf{S}^m)  \mathscr{Y}^{(k_1)q_1}(\mathbf{S}^i)
\right]  \nonumber \\
 &= (2S^m+1)(2S^i+1) \bigl[ \delta_{ \tilde{m} m} \delta_{ \tilde{j} i}  \delta_{ \tilde{k}' k'} \delta_{ \tilde{q}' q'} \delta_{\tilde{q}_2 q_1 }  \nonumber  \\
 & \quad + \delta_{ \tilde{j} m } \delta_{ \tilde{m} i}  \delta_{ 1 k'} \delta_{ \tilde{q}_2 q'}  \delta_{\tilde{k}' 1}  \delta_{ \tilde{q}' q_1} \bigr]
  \label{eq:trace2}
\end{align}
\begin{align}
& \mbox{Tr} \left[ \mathscr{Y}^{(\tilde{k}')\tilde{q}'}(\mathbf{S}^{\tilde{m}})^\dagger \mathscr{Y}^{(k_1)\tilde{q}_1}(\mathbf{S}^{\tilde{i}})^\dagger  
\mathscr{Y}^{(k')q'}(\mathbf{S}^m)  \mathscr{Y}^{(k_2)q_2}(\mathbf{S}^j) 
\right]  \nonumber \\
 &= (2S^m+1)(2S^j+1) \bigl[ \delta_{ \tilde{m} m} \delta_{ \tilde{i} j}  \delta_{ \tilde{k}' k'} \delta_{ \tilde{q}' q'} \delta_{\tilde{q}_1 q_2 }  \nonumber \\
 & \quad + \delta_{ \tilde{i} m } \delta_{ \tilde{m} j }  \delta_{ 1 k'} \delta_{ \tilde{q}_1 q'}  \delta_{\tilde{k}' 1}  \delta_{ \tilde{q}' q_2 } \bigr]
  \label{eq:trace3}
  \end{align}
\begin{align}
& \mbox{Tr} \left[ \mathscr{Y}^{(\tilde{k}')\tilde{q}'}(\mathbf{S}^{\tilde{m}})^\dagger \mathscr{Y}^{(k_1)\tilde{q}_1}(\mathbf{S}^{\tilde{i}})^\dagger  
\mathscr{Y}^{(k')q'}(\mathbf{S}^m)  \mathscr{Y}^{(k_1)q_1}(\mathbf{S}^i) \right]   \nonumber  \\
 & = (2S^m+1)(2S^i+1) \bigl[ \delta_{ \tilde{m} m} \delta_{ \tilde{i} i}  \delta_{ \tilde{k}' k'} \delta_{ \tilde{q}' q'} \delta_{\tilde{q}_1 q_1 }  \nonumber  \\
 & \quad + \delta_{ \tilde{i} m } \delta_{ \tilde{m} i }  \delta_{ 1 k'} \delta_{ \tilde{q}_1 q'}  \delta_{\tilde{k}' 1}  \delta_{ \tilde{q}' q_1 } \bigr].
  \label{eq:trace4}
  \end{align}
Collecting all the terms, and restricting the various summations according to the various Kronecker delta functions, we get the overall dissipation rate.  This multi-index expression, which involves substituting Eqs.~(\ref{eq:trace1}), (\ref{eq:trace2}), (\ref{eq:trace3}), (\ref{eq:trace4}), (\ref{eq:LL}), (\ref{eq:LR}) into (\ref{eq:IP}), will not be provided here, as it is neither informative, nor particularly illuminating.  We only need to highlight some key features.
We see that the dissipation rate of the one-body term, $W(\mathscr{Y}^{(1)q}(\mathbf{S}^m))$, is proportional to bath autocorrelation functions of the form $\langle A_{ \tilde{k} \tilde{q} }( \mathbf{S}^{\tilde{i}}, \mathbf{S}^{\tilde{j}} )(0)^\dagger,  A_{kq}( \mathbf{S}^i, \mathbf{S}^j)(\tau) \rangle_B$, which according to Abragam~\cite{abragam1961principles}  must be taken as thermal averages, i.e. $\langle AA(t) \rangle_B \equiv \mbox{tr}[\rho AA(t) ]$, where $\rho(\mathcal{H}_B)=\exp(-\beta_L \mathcal{H}_B)/\mbox{Tr}[\exp(-\beta_L \mathcal{H}_B)]$ is the Boltzmann density matrix and $\mbox{tr}$ denotes the partial trace over the bath degrees of freedom.

Assuming that spins are all of the same type ($S^i=S$), two types of terms arise.   The first is:
\begin{widetext}
\begin{align*}
 \Gamma_1^{ k q \tilde{k} \tilde{q} }(q\omega)   = & \int_0^\infty d\tau \sum_j \sum_{ \tilde{i} < j } \sum_{ i < j} \langle A_{ \tilde{k} \tilde{q} } ( \mathbf{S}^{\tilde{i}}, \mathbf{S}^{j})^\dagger(0) A_{kq}(\mathbf{S}^i,\mathbf{S}^j)(\tau) \rangle_B e^{-i\omega q \tau}  \\
 =& \int_0^\infty d\tau \int_{\mathbb{R}^3} d^3\mathbf{r}_1 \int_{ |\mathbf{r}_3-\mathbf{r_1} | > \epsilon } d^3\mathbf{r}_2 \int_{ |\mathbf{r}_3-\mathbf{r_1} | > \epsilon } d^3\mathbf{r}_3 \langle A_{ \tilde{k} \tilde{q} } ( \mathbf{r}_2, \mathbf{r}_1)^\dagger(0) A_{kq}( \mathbf{r}_3, \mathbf{r}_1 )(\tau)  \rangle_B e^{-i\omega q \tau}  \\
  =& \int_0^\infty d\tau \int_{  |\mathbf{r}_{31}|>\epsilon  } d^3\mathbf{r}_{31} \int_{ |\mathbf{r}_{21}|>\epsilon } d^3\mathbf{r}_{21} \int_{\mathbb{R}^3} d^3\mathbf{r}_1 \langle A_{ \tilde{k} \tilde{q} } ( \mathbf{r}_{21} )^\dagger(0) A_{kq}( \mathbf{r}_{31} )(\tau)  \rangle_B e^{-i\omega q \tau}  \\
  =& \int_0^\infty d\tau \int_{\mathbb{R}^3} d^3\mathbf{r}_{1} \langle \overline{ A_{ \tilde{k} \tilde{q} } ( \mathbf{r}_{21} )^\dagger(0) } \cdot \overline{ A_{kq}( \mathbf{r}_{31} )(\tau)  } \rangle_B e^{-i\omega q \tau}  
 \end{align*}
\end{widetext}
where $\epsilon>0$, $\mathbf{r}_{21} \equiv \mathbf{r}_2-\mathbf{r}_1$, $\mathbf{r}_{31} \equiv \mathbf{r}_3-\mathbf{r}_1$ and the overline denotes a volume integral over $\mathbf{r}_{21}$ or $\mathbf{r}_{31}$ (as seen from the point $\mathbf{r}_{1}$, but excluding it).  All integrals, including the one over $\mathbb{R}^3$, cover only the volume of interest (E-qubit).  In the third equality we assumed spatial homogeneity (distribution of spins is a stationary random field). This expression is the single-sided temporal Fourier transform of the volume averaged two-point spatio-temporal autocorrelation function of the spin-bath coupling.

Assuming spatial homogeneity the second type of term is of the form: 
\begin{widetext}
\begin{align*}
\Gamma^{ k q \tilde{k} \tilde{q} }_{2,m,\tilde{m}} (q\omega)  = & \int_0^\infty  d\tau \sum_{ \tilde{i} < m } \sum_{ i < \tilde{m} } \langle A_{ \tilde{k} \tilde{q} } ( \mathbf{S}^{\tilde{i}}, \mathbf{S}^m )^\dagger(0) A_{kq}(\mathbf{S}^i, \mathbf{S}^{ \tilde{m} } )(\tau) \rangle_B e^{-i\omega q \tau}  \\
=&  \int_0^\infty  d\tau \int_{  |\mathbf{r}_1-\mathbf{r}_m | > \epsilon } d^3 \mathbf{r}_1 \int_{  |\mathbf{r}_2-\mathbf{r}_{\tilde{m} } | > \epsilon } d^3\mathbf{r}_2  \langle A_{ \tilde{k} \tilde{q} } ( \mathbf{r}_1, \mathbf{r}_m )^\dagger(0) A_{kq}( \mathbf{r}_2, \mathbf{r}_{ \tilde{m}} )(\tau)  \rangle_B e^{-i\omega q \tau}   \\
=&  \int_0^\infty d\tau \int_{  |\mathbf{r}_{1m}|>\epsilon  } d^3 \mathbf{r}_{1m} \int_{  |\mathbf{r}_{2\tilde{m}}|>\epsilon  } d^3\mathbf{r}_{2\tilde{m}}   \langle A_{ \tilde{k} \tilde{q} } ( \mathbf{r}_{1 m} )^\dagger(0) A_{kq}( \mathbf{r}_{2 \tilde{m}} )(\tau)  \rangle_B e^{-i\omega q \tau}   \\
  =& \int_0^\infty d\tau  \langle \overline{ A_{ \tilde{k} \tilde{q} } ( \mathbf{r}_{1 m} )^\dagger(0) }  \cdot \overline{ A_{kq}( \mathbf{r}_{2 \tilde{m}} ) (\tau) } \rangle_B e^{-i\omega q \tau},
\end{align*}
\end{widetext}
where $\epsilon>0$, $\mathbf{r}_{1m} \equiv \mathbf{r}_1-\mathbf{r}_m$, $\mathbf{r}_{2\tilde{m}} \equiv \mathbf{r}_2-\mathbf{r}_{\tilde{m}}$ and the overline denotes a spatial autocorrelation function for the spin-bath interaction (as seen from $\mathbf{r}_{1m}$ and $\mathbf{r}_{2\tilde{m}}$ but excluding those points).   This dissipation rate describes how spatial autocorrelations functions of the microscopic field are temporally correlated.  The spectral density functions $\Gamma^{ k q \tilde{k} \tilde{q} }_{1} (q\omega) $ and $\Gamma^{ k q \tilde{k} \tilde{q} }_{2,m,\tilde{m}} (q\omega) $ combine to give the overall dissipation rate involving the states $|F_m\rangle\rangle$ and $|F_{\tilde{m}}\rangle\rangle$.  The remaining summations and indices are geometrical factors related to the selection rules for coupling of angular momenta.

\subsubsection{Many-Body States}

Correlated many-body states can arise due to quantum state preparation or via coherence evolution in the presence of residual couplings.  Their dissipation rate is much faster. 
Consider the $n$-spin state
$$ F = \mathscr{Y}^{(1)q_1}(\mathbf{S}^1) \mathscr{Y}^{(1)q_2}(\mathbf{S}^2) \dots \mathscr{Y}^{(1)q_n}(\mathbf{S}^n)   $$
where $q_i \in \{ -1,0,1\}$. The idea here is that when computing $\mathscr{L}F=[ \mathcal{H}, F]$, and making use of the commutator rule
$$ [ \mathcal{H}, A_1 \dots A_n] = [ \mathcal{H}, A_1] A_2 \dots A_n $$
$$ + A_1 [ \mathcal{H}, A_2] A_3 \dots A_n + \dots + A_1 \dots A_{n-1} [ \mathcal{H}, A_n], $$
which yields $n$ terms, there will be $n^2$ terms in total 
when computing the inner product $\left( \mathscr{L}^*(0) F, e^{\mathscr{L}_B \tau } \mathscr{L}^*(\tau) F \right)$ since $\mathscr{L} F$ appears twice. And because each term contributes a dissipation rate of the same magnitude as the ones computed in the previous section (for single-spin term), the overall dissipation rate, $W(\mathscr{Y}^{(1)q_1}(\mathbf{S}^1) \mathscr{Y}^{(1)q_2}(\mathbf{S}^2) \dots \mathscr{Y}^{(1)q_n}(\mathbf{S}^n))$, which is a sum of $n^2$ such terms, decays $n^2$ times faster than in the case of the single-body state.


From this we conclude that entangled many-body states are more vulnerable to decoherence than uncoupled single-spin states.
Their dissipation rates can in principle be computed exactly from the above analysis, if the bath two-body spatio-temporal autocorrelation functions are known.

\typeout{}
\bibliography{QuditBibliography.bib}

\end{document}